# JWST/MIRI Imaging of the Warm Dust Component of the $\epsilon$ Eridani Debris Disk

Schuyler G. Wolff[1], András Gáspár[1], George Rieke[1], Jarron M. Leisenring[1], Antranik A. Sefilian[1], Marie Ygouf[2], and Jorge Llop-Sayson[2]

[1] Steward Observatory and the Department of Astronomy, The University of Arizona, 933 N Cherry Ave., Tucson, AZ 85721, USA; sgwolff@arizona.edu

[2] Jet Propulsion Laboratory, California Institute of Technology, Pasadena, CA 91109, USA

## Abstract

We present JWST/MIRI observations of the debris disk surrounding the nearby solar analog $\epsilon$ Eridani obtained as part of the Archetypal Debris Disk Guaranteed Time Observation program. Multiwavelength images from 15, 18, 21, and 25.5 $\mu$m show a smooth dust distribution with no evidence of sculpting by massive planets outside of 5 au. Maps of the color temperature and opacity constrain the dust properties, while radiative transfer modeling of a warm dust component traces the interaction between the debris disk and $\epsilon$ Eridani b (∼3.5 au). Dynamical and collisional modeling further shows that the disk morphology is dominated by dust produced in the outer planetesimal belt (∼70 au) moving inward via stellar wind drag. We confirm the presence of a disk interior to the $\epsilon$ Eri b orbit first detected from mid-IR interferometry. Drag-dominated inner disk regions have also been observed around Vega and Fomalhaut, hinting at the diversity of asteroid belt analogs.

## 1. Introduction

$\epsilon$ Eridani (hereafter $\epsilon$ Eri) is a relatively young (200–800 Myr; E. E. Mamajek & L. A. Hillenbrand 2008; C. L. Sahlholdt et al. 2019), K2 spectral-type star with a mass of 0.82 $M_\odot$ (E. K. Baines & J. T. Armstrong 2012) that hosts both planet(s) and a debris disk. At a distance of only 3.2 pc, the system has been beguiling astronomers since the first detection of its circumstellar disk with IRAS (H. H. Aumann 1985). It is the closest prominent debris disk, and hence offers unique opportunities for study at high spatial resolution. The disk has been extensively observed across all wavelengths. The emerging picture of the disk structure is complex. In the outer regions, Atacama Large Millimeter/submillimeter Array (ALMA) imaging shows a remarkably narrow outer ring centered at 69 au with a width of 12 ± 1 au and an inclination with respect to the sky plane of 33°.7 ± 0°.5 (M. Booth et al. 2017, 2023). This belt has the thermal equivalent distance of the Kuiper Belt, and can be considered an analog. In the innermost regions, a strong exozodiacal dust detection (∼300 zodis) is seen with nulling interferometry at 10 $\mu$m (HOSTS survey; S. Ertel et al. 2020). The exozodi profile appears to be flat out to ∼6 au (private communication with the HOSTS team).

The structure of the intermediate disk regions has been more difficult to interpret. Partially resolved imaging and spectral information are consistent both with an intermediate belt(s) located somewhere in the 4–50 au region (D. Backman et al. 2009; J. S. Greaves et al. 2014) or with a population of smaller dust particles produced in the outer ring and pulled in via drag forces (M. Reidemeister et al. 2011). K. Y. L. Su et al. (2017) reanalyzed the existing observations in combination with a SOFIA 35 $\mu$m radial profile and found that a model with intermediate disk flux dominated by dragged-in grains is disfavored. The M. Reidemeister et al. (2011) model underpredicts the 35 $\mu$m disk flux within 4″ (12 au), and overpredicts it outside of 6″ (20 au). We note, however, that the addition of planet(s) complicates this interpretation.

$\epsilon$ Eridani is also host to at least one planet. The first detection of $\epsilon$ Eridani b was by A. P. Hatzes et al. (2000) via radial velocity (RV). The precise orbital parameters of $\epsilon$ Eri b are an active area of study. Results from J. Llop-Sayson et al. (2021), F. Feng et al. (2023), and D. Bisht & H. R. A. Jones (2024) all agree reasonably well. For example, F. Feng et al. (2023) combine RV and several astrometric data sets including (multiepoch) GAIA observations to give a best-fit mass of $0.76^{+0.14}_{-0.11}$ $M_{\rm Jup}$, semimajor axis of $3.53^{+0.06}_{-0.06}$ au, an inclination of $130.60^{+9.53}_{-12.62}$ degrees, and an eccentricity of $0.26^{+0.04}_{-0.04}$. Most recently, a similar analysis[3] by W. Thompson et al. (2025) reports posterior 68% credible intervals for a planet mass of $0.98^{+0.10}_{-0.09}$ $M_{\rm Jup}$, a semimajor axis of $3.53^{+0.04}_{-0.04}$ au, an inclination of $41^{+6}_{-5}$ degrees, and an eccentricity of $0.06^{+0.06}_{-0.04}$.

There may be an additional planet in the outer system. A narrow ring of thermal emission from the cold outer disk component at millimeter wavelengths has been detected with ALMA (M. Booth et al. 2017, 2023) centered at 69 au. Clumpy structures observed in the ring are posited to be caused by mean-motion resonances with a migrating planet (at a separation of ∼40 au) trapping planetesimals in the 2:1 mean-motion resonance.

Recent attempts to pin down the intermediate disk structure have been ineffective. In S. G. Wolff et al. (2023), we reported on a deep, but unsuccessful, search for scattered light from the $\epsilon$ Eri debris system using the Hubble Space Telescope (HST). Modeling of this nondetection in combination with information on the disk spectrum available in the literature, we found that (1) a minimum grain size of ∼2 $\mu$m in the outer disk is consistent with the spectral energy distribution (SED), and (2) there is evidence for grain-size stratification with the Spitzer Infrared Spectrograph (IRS) spectrum preferring

[3] The key difference between F. Feng et al. (2023) and W. Thompson et al. (2025) is the treatment of the (correlated) uncertainties across the GAIA and Hipparcos data sets.

**Table 1**
JWST/MIRI Observations of the $\epsilon$ Eridani System (2024 February 12)

| Target | Filter/Subarray | PA[a] (deg) | $N_{group}$ | $N_{int}$ | Dither | Time (s) |
|---|---|---|---|---|---|---|
| Background #1 | F1500W/SUB256 | 77.05 | 5 | 150 | 4pt Ext | 1077 |
| Background #1 | F1800W/BSKY | 77.06 | 5 | 48 | 4pt Ext[b] | 993 |
| Background #1 | F2100W/BSKY | 77.06 | 5 | 48 | 4pt Ext[b] | 993 |
| Background #1 | F2550W/BSKY | 77.06 | 5 | 52 | 4pt Ext[b] | 1076 |
| $\delta$ Eri (PSF Ref.) | F2550W/BSKY | 77.09 | 5 | 52 | 4pt Ext[b] | 1076 |
| $\delta$ Eri (PSF Ref.) | F2100W/BSKY | 77.09 | 5 | 48 | 4pt Ext[b] | 993 |
| $\delta$ Eri (PSF Ref.) | F1800W/BSKY | 77.09 | 5 | 48 | 4pt Ext[b] | 993 |
| $\delta$ Eri (PSF Ref.) | F1500W/SUB256 | 77.12 | 5 | 150 | 4pt Ext | 1077 |
| Background #2 | F1500W/SUB256 | 77.64 | 5 | 150 | 4pt Ext | 1077 |
| Background #2 | F1800W/BSKY | 77.66 | 5 | 48 | 4pt Ext[b] | 993 |
| Background #2 | F2100W/BSKY | 77.66 | 5 | 48 | 4pt Ext[b] | 993 |
| Background #2 | F2550W/BSKY | 77.66 | 5 | 52 | 4pt Ext[b] | 1076 |
| $\epsilon$ Eri (Rot. 1) | F2550W/BSKY | 72.83 | 5 | 52 | 4pt Ext[b] | 1076 |
| $\epsilon$ Eri (Rot. 1) | F2100W/BSKY | 72.83 | 5 | 48 | 4pt Ext[b] | 993 |
| $\epsilon$ Eri (Rot. 1) | F1800W/BSKY | 72.83 | 5 | 48 | 4pt Ext[b] | 993 |
| $\epsilon$ Eri (Rot. 1) | F1500W/SUB256 | 72.83 | 5 | 150 | 4pt Ext | 1077 |
| $\epsilon$ Eri (Rot. 2) | F1500W/SUB256 | 82.84 | 5 | 150 | 4pt Ext | 1077 |
| $\epsilon$ Eri (Rot. 2) | F2550W/BSKY | 82.83 | 5 | 52 | 4pt Ext[b] | 1076 |
| $\epsilon$ Eri (Rot. 2) | F2100W/BSKY | 82.83 | 5 | 48 | 4pt Ext[b] | 993 |
| $\epsilon$ Eri (Rot. 2) | F1800W/BSKY | 82.83 | 5 | 48 | 4pt Ext[b] | 993 |

**Notes.**
[a] On-sky position angle of observed aperture.
[b] The BRIGHTSKY subarray four-point dither patterns used the #6 starting set with a single set.

smaller grains in the inner disk regions. Subsequent attempts to detect the disk in scattered light with the HST Space Telescope Imaging Spectrograph subarray (P. M. S. Krishnanth et al. 2024) and with long-term SPHERE polarimetric monitoring (C. Tschudi et al. 2024) were also unsuccessful.

The lack of small dust grains, particularly in the regions near the outer planetesimal belt, is unusual. Small dust grains collisionally replenished from the 70 au planetesimal belt (J. S. Dohnanyi 1969) can be removed from the system by a combination of radiation pressure and drag forces (both Poynting–Robertson and corpuscular stellar wind drag; T. Löhne et al. 2008). J. A. Arnold et al. (2019) predict that all grain sizes will be stable against blowout for astronomical silicates ,and we do not expect radiation forces to be the dominant source of removal. $\epsilon$ Eri is known to have a strong stellar wind (SW) with a mass-loss rate 16 times solar (K. G. Kislyakova et al. 2024), likely dominating the forces acting on the $\sim$1 $\mu$m dust particles.

With JWST/MIRI, we are able to probe thermal emission from dust of a few to a few tens of microns in size where the star/disk contrast is more favorable and at a spatial resolution never before achieved in the mid-IR. We present multi-wavelength MIRI imaging observations in Section 2 and the data reduction and image processing in Section 3. Analysis of the architecture and photometry of the disk are presented in Section 4. The long-term dynamical and collisional history for $\epsilon$ Eri is modeled in Section 5. The disk dust properties, interaction with $\epsilon$ Eri b, and drag-dominated inner region are discussed in Section 6, and conclusions are given in Section 7.

## 2. JWST Observations

We observed the $\epsilon$ Eri system with JWST/MIRI on 2024 February 12 as part of Guaranteed Time Observation (GTO) program 1193, alongside Fomalhaut (A. Gáspár et al. 2023) and Vega (K. Y. L. Su et al. 2024). Prior experience with the Fomalhaut and Vega data sets led us to reconfigure the $\epsilon$ Eri observations presented here, removing the coronagraphy components and focusing solely on direct-imaging observations in similar filter bandpasses. While the coronagraphic filters allow longer integration ramps, ideal for the imaging of faint point sources, they also introduce point-spread function (PSF) and imaging artifacts and limit the field of view (FOV) of the observations. In addition, the F2300C Lyot coronagraph severely limits the inner working angle (IWA) of the observations due to the opaque occulting mask. Direct imaging, on the other hand, eliminates these limitations and, given the remarkable stability of the observatory optical telescope assembly at mid-IR wavelengths, enables precise subtractions of the stellar PSFs. Direct imaging's primary disadvantage stems from the saturation of the stellar core for bright sources such as $\epsilon$ Eri. However, the saturated region is smaller than the one blocked by the F2300C Lyot mask and yields a similarly useful IWA for this system at 15.0 $\mu$m as the F1550C coronagraph. While longer integration ramps may increase the depth for faint features in the coronagraph's background-limited regime, direct imaging has been shown empirically to provide improvements when imaging extended thermal emission around stars. The observations presented here were therefore imaged at at 15.0 (F1500W), 18.0 (F1800W), 21.0 (F2100W), and 25.5 $\mu$m (F2550W). All the JWST data used in this paper can be found in MAST (doi:10.17909/trht-ak70).

We summarize the MIRI observations of the $\epsilon$ Eri system in Table 1. Alongside the MIRI observations, JWST/NIRCam data were also acquired to search for exoplanet companions; these will be presented in a parallel paper (J. Llop-Sayson et al. 2025) and are not discussed here. The sequence of observations was optimized to minimize latent artifacts in the MIRI detectors as well as overheads. As our observations intentionally used short ramp sampling (five groups for all wavelengths) to limit the saturation of the stellar cores, where dark corrections are not well calibrated, the PID 1193 MIRI observations used dedicated background observations not only to correct the astrophysical backgrounds and detector artifacts but also the instrumental dark levels. For this reason, the backgrounds had to be observed identically to the target and the PSF. Since the observations of the bright target and PSF leave latent patterns in the detectors, we begin the sequence by observing these dedicated backgrounds at each wavelength, guaranteeing a clean image. We observed two background regions within this program, one near the PSF and one near the science target. The reasons for this were two-fold: (1) to analyze the variations in the sky background, (2) to allow the latent signature of the PSF to decay during the ~1 hr of background observation (D. Dicken et al. 2024). During data reduction, it was discovered that the second background sequence was redundant; the two sets were nearly identical apart from latent residuals present in the background observed after $\delta$ Eridani (hereafter $\delta$ Eri). Since the presence of this persistent image has no practical use during imaging processing, we opted to discard the second background sequence. However, these observations were useful to confirm that the background remained constant at these two locations and should help inform observing strategies for future observations. Figure 1 shows these two background images and their difference.

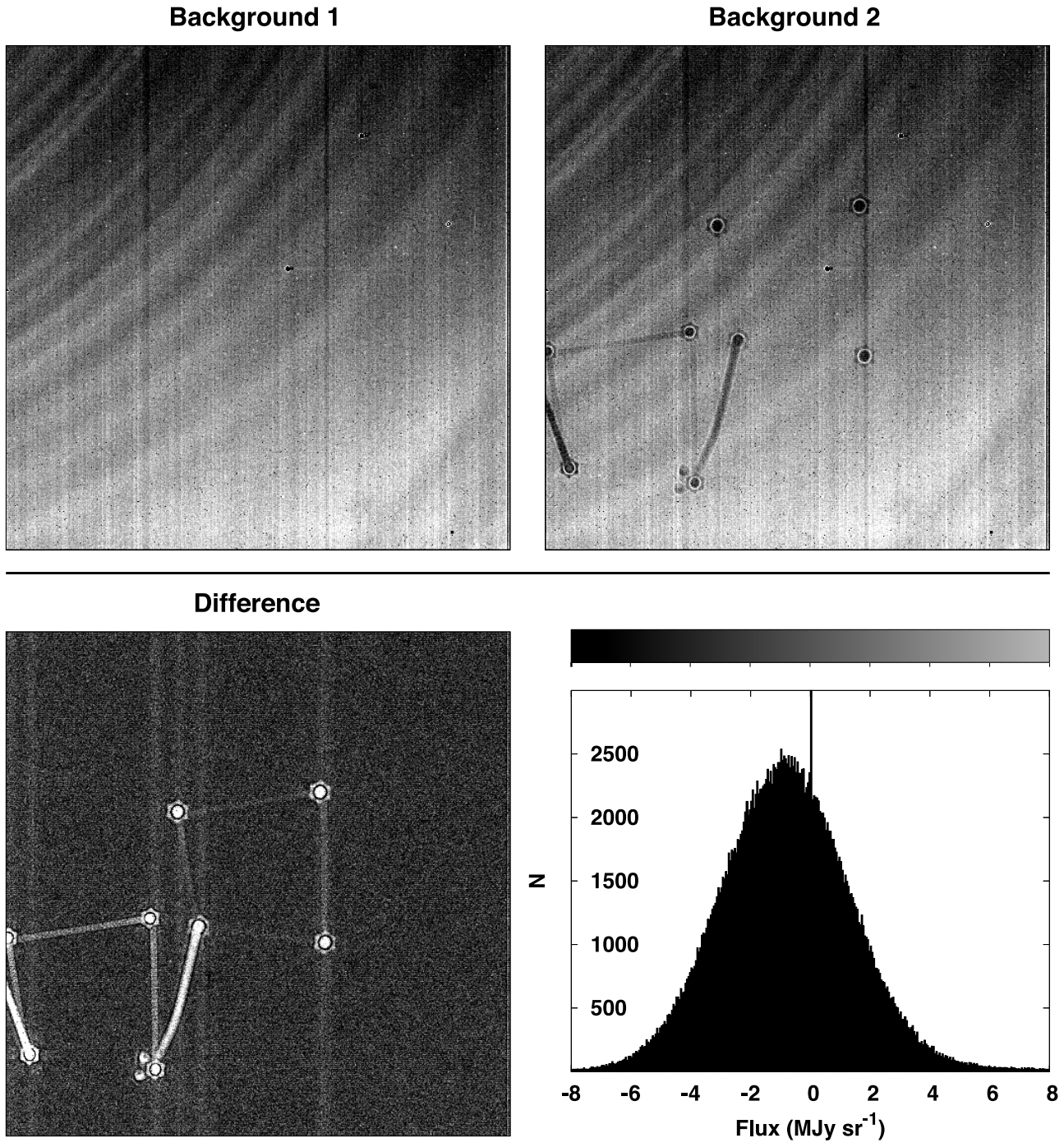

**Figure 1.** The background on blank sky at 25.5 $\mu$m, near $\delta$ Eri (Background 1) and $\epsilon$ Eri (Background 2), scaled linearly between 820 and 1000 MJy sr$^{-1}$. The latent patterns imprinted on the detector, following the observation of $\delta$ Eri, are apparent in background 2 (and in the difference image, shown in the bottom panel, scaled between −8 and 8 MJy sr$^{-1}$). The background imaged at $\epsilon$ Eri is on average 0.84 MJy sr$^{-1}$ brighter than the one at $\delta$ Eri.

Detector subarrays were chosen to reduce frame times and minimize saturation while also maximizing the imaging area around the science target where prior observations indicate the presence of disk emission. We found the SUB256 subarray for the F1500W imaging and the BRIGHTSKY subarray for the other three filters (F1800W, F2100W, F2550W) to be optimal for this strategy. For dithering, we used the four-point extended pattern, which maximizes displacement; this was initiated at starting point #6 for the BRIGHTSKY subarray. This dither pattern places the source at an ideal location to recover extended features far from the stellar core as well as to allow for the removal of latent signatures. The PSF was observed with identical dither patterns as the science target, which is critical for high-accuracy removal of the stellar PSF (see, e.g., A. Gáspár et al. 2023, K. Y. L. Su et al. 2024). We also observed the science target at two rotation angles, which also proved to be critical in recovering areas masked due to latent artifacts.

For PSF reference, we chose the nearby star $\delta$ Eri, which has been used as a reference source for $\epsilon$ Eri over the years of observing with HST (e.g., S. G. Wolff et al. 2023). The properties that make this star an ideal reference source are its similar brightness to $\epsilon$ Eri (ensuring similar noise properties in its images and that both images are similarly affected by the brighter-fatter effect), its lack of a companion, and its proximity to our science target in the sky (reducing thermal artifacts from the observatory). The reference target is also a similar spectral type to our science target—serendipitously—although this is not a requirement for long-wavelength MIRI observations (see, e.g., A. Gáspár et al. 2023).

## 3. Data Reduction and Image Processing

The data reduction and image processing steps followed the same sequence described in A. Gáspár et al. (2023) and K. Y. L. Su et al. (2024). Here we detail the basic processing steps we performed. For brevity, we only showcase the reduction of the F2550W data set, as the approach was identical at all four wavelengths.

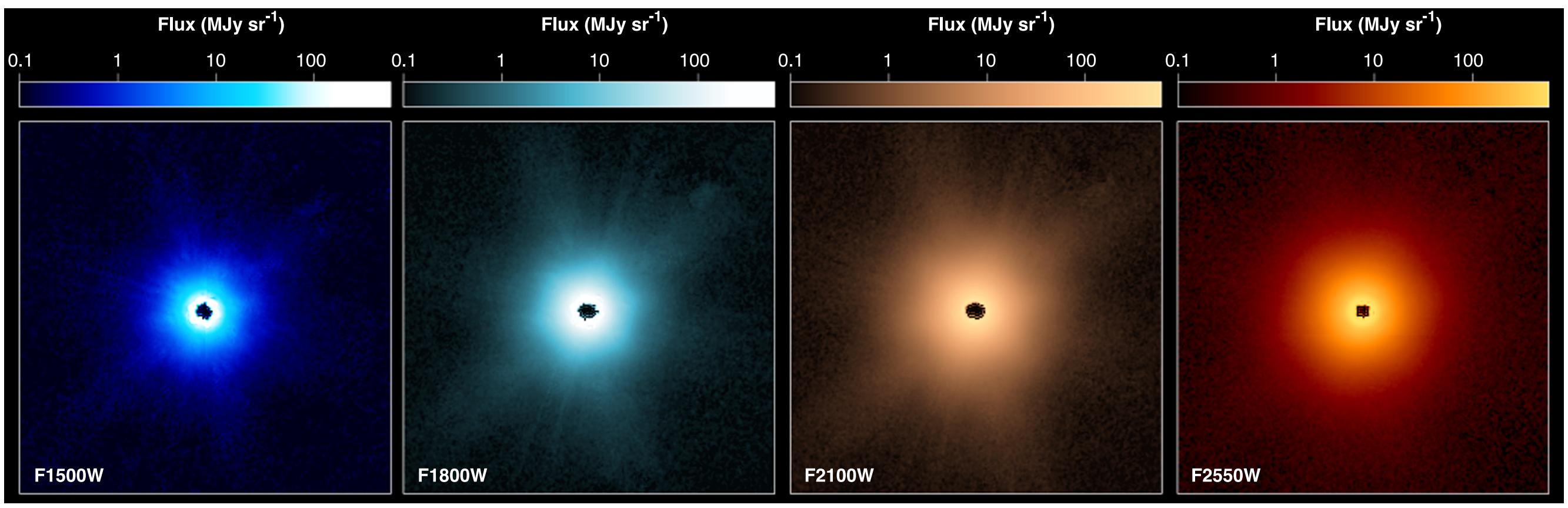

**Figure 2.** The final data products of our JWST/MIRI $\epsilon$ Eri observations at the four wavelengths of the observations. Images are oriented north up and east to the left, with a displayed FOV of 28."16 × 28."16 to highlight the bright, inner disk regions. The outer planetesimal belt (observed by ALMA at ∼70 au) is not detected and is outside the FOV of these images.

We downloaded uncalibrated (_uncal) data from the Barbara A. Mikulski Archive for Space Telescopes (MAST), allowing custom reduction steps to be applied. For the step 1 and step 2 processing, we used version 1.13.4 (released 2024 January 25) of the JWST pipeline, keeping most of the settings at their default values. The step 1 processing corrects for detector-level artifacts, such as bad pixels, nonlinear response in count rates, cosmic-ray hits, and dark current. As the dark current is not well measured for the short integration ramps used in our observing strategy, we turned this correction off and used the dedicated and identically observed background images to subtract the dark current. The pipeline also skips the reset charge decay correction, automatically, for such short integration ramps. The jump rejection threshold for cosmic rays was set to 5$\sigma$ (the default value is 4$\sigma$).

Following the stage 1 and 2 reduction steps, we performed custom image processing steps using IRAF and idp3, including background subtractions, image alignment, scaling, PSF subtraction, and row/column corrections. The first step in the image processing was to remove the background from the science and PSF reference images. While the dedicated background observations are dominated by the telescope emission, they also contain the dominant detector artifacts, such as the tree rings. Adopting the same exposure settings for our science, PSF, and background observations ensures that background subtraction also corrects for the dark current. As noted in Section 2, we used only the first set of background images acquired at the start of our observing sequence. The images taken at the four dither positions were median-averaged using a 3$\sigma$ rejection around the data median and then the results were subtracted from the science and PSF reference images. Although the background subtraction removed the majority of the artifacts present, a small level of sky offset and row/column artifacts remained. These were removed via additional processing steps, similar to those detailed in K. Y. L. Su et al. (2024) and discussed further in Appendix A.

Figure 2 shows the final disk images at all MIRI wavelengths observed in our program, while Figure 3 presents a false-color combination image of our MIRI observations. Residual diffraction spikes, most prominent at 18 and 21 $\mu$m, are indicative of the presence of marginally resolved warm disk emission in the system close to $\epsilon$ Eri.

## 4. Analysis

The final images with MIRI showcase an extended disk around $\epsilon$ Eri (see Figures 2 and 3). The disk appears to be remarkably smooth, with no signs of gaps, rings or asymmetries. It extends from the effective IWA set by the saturated region at ∼3 au (∼1″) out to the wavelength-dependent thermal noise floor, reached at ∼40 au. We do not detect the outer planetesimal belt previously seen with Herschel at far-IR (J. S. Greaves et al. 2014) or at radio wavelengths with the James Clerk Maxwell Telescope (J. S. Greaves et al. 1998), the Large Millimeter Telescope (M. Chavez-Dagostino et al. 2016), the Australia Telescope Compact Array (M. A. MacGregor et al. 2015), or ALMA (M. Booth et al. 2017, 2023).

For the background sources in the images, we cross-check against existing HST observations obtained in 2021 January (S. G. Wolff et al. 2023) for common proper motion. The source with a green color to the southeast in Figure 3 is the same source identified in M. Booth et al. (2023), though we do not detect their northwest (NW) point source. Several of the extended sources in our images are not present in HST observations and are assumed to be background galaxies.

We present the overall disk morphology in Section 4.1, and report JWST disk photometry in Section 4.2. We focus discussion on the outer disk in Section 4.3, and model the inner disk structure in Section 4.4.

### 4.1. Disk Morphology

In this section, we focus on the radial extent and color of the main disk. We first fit for the disk geometry, deproject the disk, and compute a radial profile for each image. A unique feature of the $\epsilon$ Eri data set is the radial PSF residuals. These were not observed in either the Fomalhaut or the Vega data sets (A. Gáspár et al. 2023; K. Y. L. Su et al. 2024), which used the same observational setup (at 25.5 $\mu$m). The spokes are at the correct orientation to match the six JWST/MIRI PSF diffraction spikes, but are not as sharp as would be expected for a point source. The point-source diffraction pattern for MIRI exhibits a discrete, radially evolving speckle pattern not seen in these residuals (compare to the stellar PSF in Figure 19 in Appendix A). Instead, we hypothesize these result from warm dust interior to the effective IWA of our data set (set by the saturated region close to the central star for these unocculted images). We investigate this further in Section 4.4.

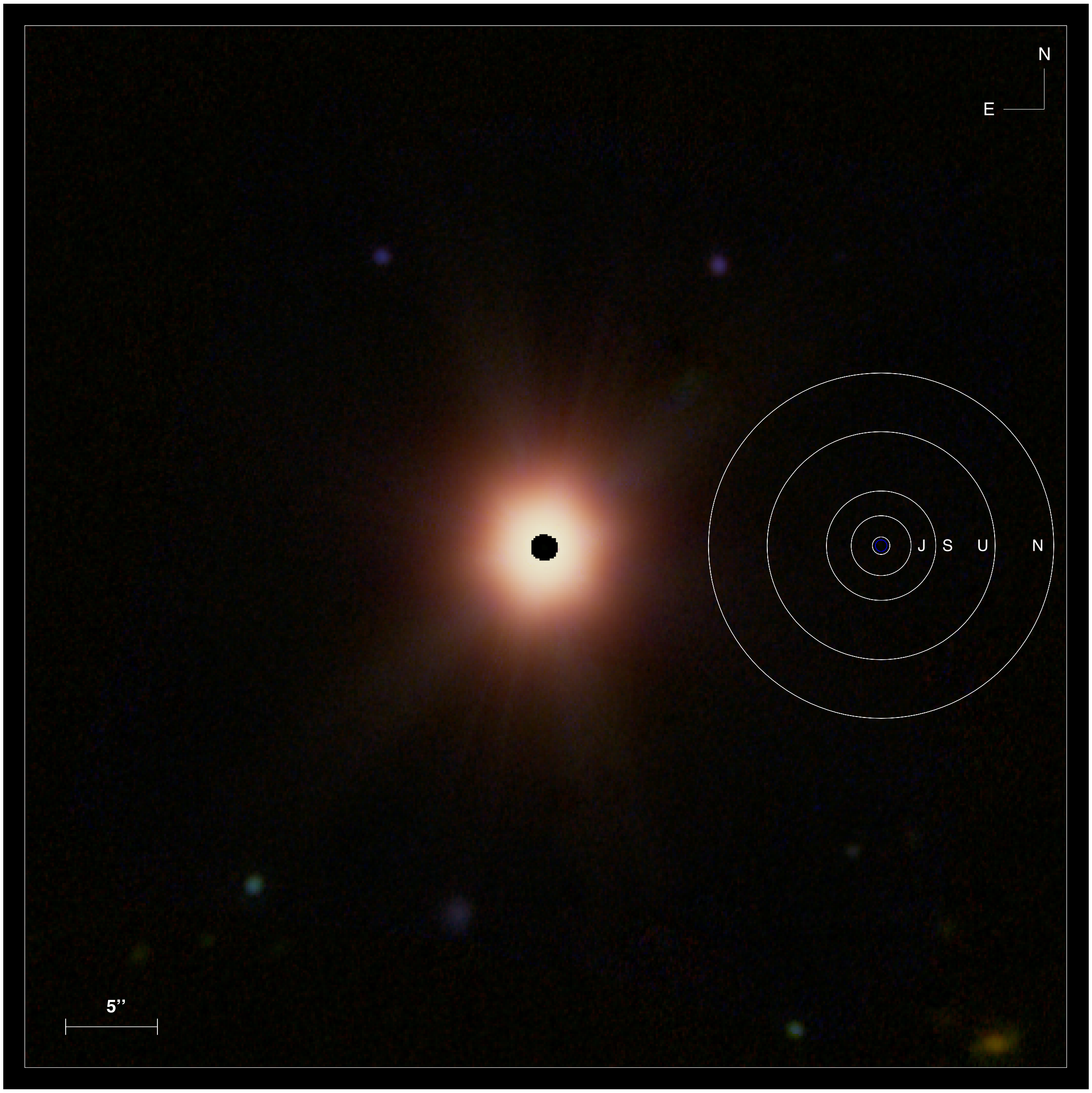

**Figure 3.** False-color combination image, using all four MIRI imaging wavelength bands. For scale, we display the spatial extent of the giant planets in the solar system, highlighting that we are spatially resolving a relatively extended disk component in the $\epsilon$ Eri system.

To determine the orbital parameters of the disk, we fit an ellipse to the image contours for each filter separately, using a contour surface-brightness value matched to a semimajor axis of $\sim$11 au along the major axis of the disk (12.0, 31.4, 60.5, and 103 MJy sr$^{-1}$ for F1500W, F1800W, F2100W, and F2550W, respectively). This location provides a high signal-to-noise ratio disk detection, but is far enough from the central star that the impacts of residual PSF structure are mitigated. Each image is subdivided into wedges with a width of $2\overset{\circ}{.}5$. For each wedge, the $x$, $y$ position that most closely matches the contour value is determined. These ($x$, $y$) coordinates are then fit with an ellipse using the Kowalsky deprojection technique for binary systems (W. M. Smart 1930) to translate the best-fit ellipse parameters into the orbital parameters for the dust particles.

We perform separate Markov Chain Monte Carlo (MCMC) fits for each filter using the `emcee` package (D. Foreman-Mackey et al. 2013) with 32 walkers and 5000 iterations. Autocorrelation times were 60–70 across all parameters and all MCMC runs. The values for the deprojected semimajor axis, the eccentricity ($e$), the inclination ($i$), the position angle (PA) measured anticlockwise from north, and the argument of periastron ($\omega$) are provided in Table 2. The given values for each parameter represent the median with errors estimated using the 16th and 84th quantiles. The orbital parameters agree across all four filters, with the largest discrepancy being in the argument of periastron, which we expect to be poorly defined for the ensemble of dust particles in an annulus. The results of the MCMC run for the F2550W filter are shown in Figure 4. This filter has the least contamination from PSF artifacts, and we adopt the best-fit values for the disk parameters optimized at 25.5 $\mu$m.

These values are in good agreement with the parameters of the narrower ALMA ring ($i = 33\overset{\circ}{.}7 \pm 0\overset{\circ}{.}5$ and PA $= -1\overset{\circ}{.}1 \pm 1\overset{\circ}{.}0$; M. Booth et al. 2023), albeit with larger uncertainties. We note that M. Booth et al. (2023) did not include a stellocentric offset in their models and thus do not include an eccentricity estimate. The best-fit eccentricity measured here corresponds to a stellocentric offset of 2.6 pixels and is consistent with an eccentricity of 0.0 within $2\sigma$. To test the impact of the differences in the best-fit orbital values, we computed deprojected radial profiles using both orbital solutions and confirmed that they agree within the uncertainties.

**Table 2**
The MCMC-fitted Orbital Parameters of the Disk

| Disk Parameter | F1500W | F1800W | F2100W | F2550W | Combined |
|---|---|---|---|---|---|
| Semimajor axis (au) | $10.9^{+0.7}_{-0.9}$ | $10.9^{+0.7}_{-0.9}$ | $10.9^{+0.7}_{-0.9}$ | $11.0^{+0.7}_{-0.9}$ | $10.9^{+0.7}_{-0.9}$ |
| Inclination, $i$ (deg) | $34.5^{+9.0}_{-7.8}$ | $33.1^{+8.9}_{-7.9}$ | $32.9^{+9.1}_{-7.9}$ | $34.5^{+9.3}_{-7.7}$ | $33.8^{+9.1}_{-7.8}$ |
| Eccentricity, $e$ | $0.08^{+0.04}_{-0.05}$ | $0.09^{+0.04}_{-0.05}$ | $0.08^{+0.04}_{-0.05}$ | $0.09^{+0.04}_{-0.05}$ | $0.09^{+0.04}_{-0.05}$ |
| PA (deg) | $-10.4^{+23.9}_{-23.0}$ | $-7.4^{+24.5}_{-28.8}$ | $-7.6^{+26.3}_{-28.5}$ | $-8.6^{+22.7}_{-26.5}$ | $-8.5^{+24.4}_{-26.8}$ |
| $\omega$ (deg) | $333^{+54}_{-40}$ | $324^{+50}_{-36}$ | $326^{+56}_{-38}$ | $312^{+53}_{-30}$ | $324^{+53}_{-36}$ |

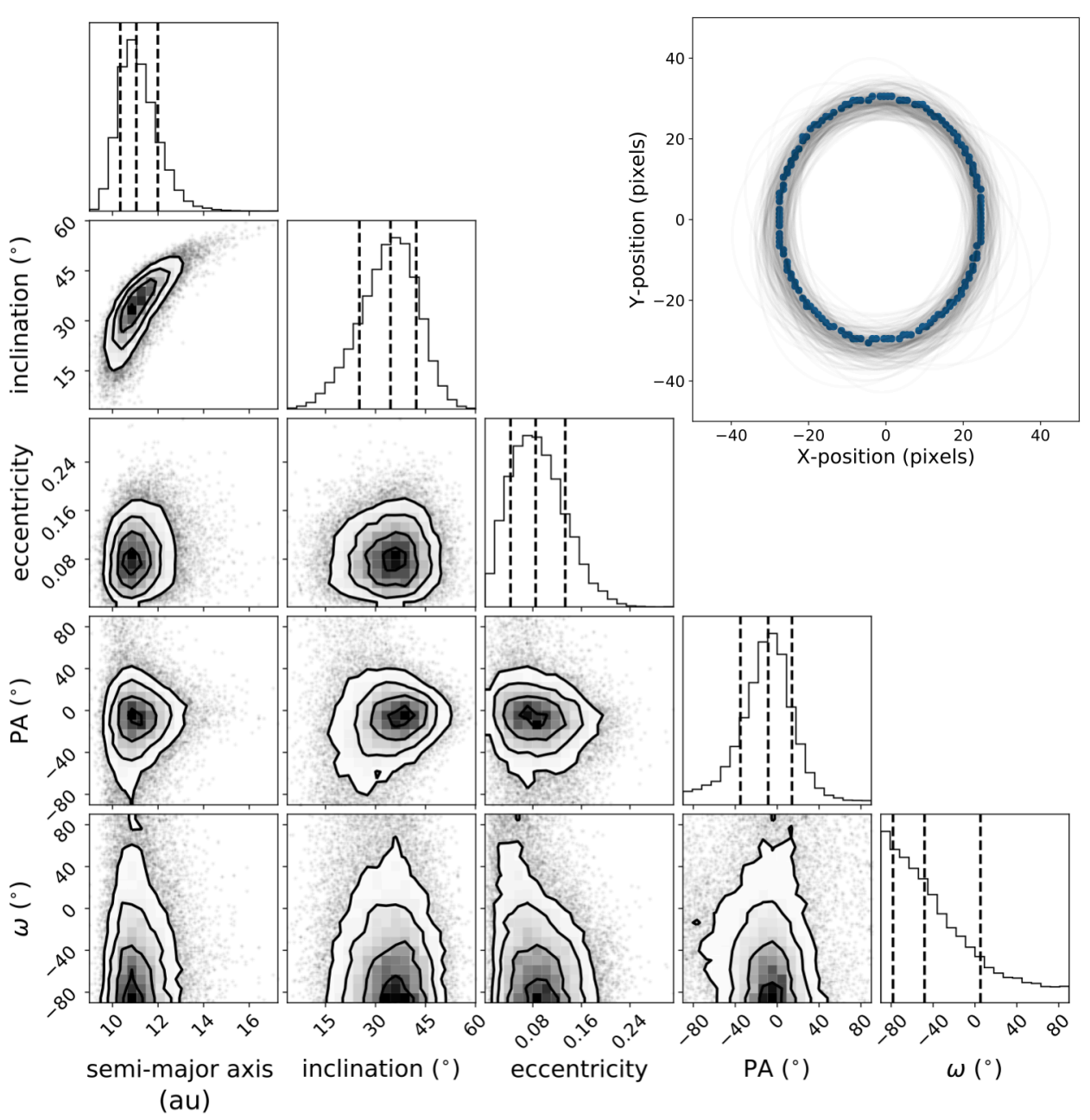

**Figure 4.** Results of the ellipse fitting to the 60.5 MJy sr$^{-1}$ contour of the F2550W image of $\epsilon$ Eri. The MCMC results are well constrained for all parameters except the argument of periastron ($\omega$), as expected. The upper-right corner shows the derived ($x$, $y$) coordinates for the contour value (blue) alongside a sample of 250 models drawn from the chain. The 16%, 50%, and 84% quantiles are shown with dashed lines and listed in Table 2.

We use the best-fit ellipse parameters to deproject the disk and compute its azimuthally averaged surface-brightness radial profiles, as shown in Figure 5. For the radial profiles, we mask the residual diffraction spikes by excluding any pixels within 5° wedges of the three diffraction spike orientations for both telescope roll angles. The radial profile is computed as the median in logarithmically spaced radial bins, with the uncertainty given by the standard deviation within each annulus.

The disk brightness increases with increasing wavelength, as expected. The disk surface brightness appears to extend smoothly out to ∼40 au. Outside of this, confusion with the increased thermal background noise makes the exact surface-brightness slope difficult to determine. Some disk contributions to the radial profiles are observed to extend past the planetesimal belt observed with ALMA (centered at 69 au) out to ∼100 au. There is a slight enhancement in surface brightness just interior to the planetesimal ring, most prominent at 15 $\mu$m. We would expect cold dust at these radii to emit more efficiently at longer wavelengths, though the decreased thermal background at the shorter wavelengths may also be playing a role. Future observations optimized for the outer disk region are needed to investigate grain production in the outer belt.

We fit power laws to the radial profiles, and find that the power-law slope from 6 to 30 au[4] decreases with increasing wavelength with a best-fit slope of −2.9 at 25.5 $\mu$m and −3.4 at 15 $\mu$m.

### 4.2. Disk Photometry

We compare the integrated disk flux at the four MIRI passbands to the flux values available in the literature. This is challenging, since the spatial extent of the disk probed by various unresolved observations differs significantly. An effort was made in S. G. Wolff et al. (2023) to decouple the flux contributions from the inner disk region to compare with the photosphere-corrected IRS spectra. Figure 6 compares those results to the MIRI observations reported here. For each MIRI image, we use the deprojected images to sum up the flux from the saturation radius of ∼0.″7 (2.3 au) to a radius of 10″. This does not include flux contributions from the saturated core (dominated by the photospheric emission) but does include contributions from the inner disk to the PSF residual spokes. The MIRI fluxes are generally consistent to within the 2% uncertainty in the IRS spectrum caused by the photospheric subtraction, though the MIRI results are systematically lower than the IRS results at longer wavelengths, with the largest discrepancy at 21 $\mu$m. This could be the result of a slight oversubtraction of the stellar flux using the adopted PSF scale factors (see Appendix A), which increase with wavelength. We adopt conservative systematic uncertainties of 10% for the MIRI disk photometry.

As a final test of the flux calibration for these MIRI observations, we compare the MIRI 25.5 $\mu$m radial profile with the lower-resolution radial profiles provided by MIPS at 24 $\mu$m and SOFIA at 35 $\mu$m (from K. Y. L. Su et al. 2017) in Figure 7. The radial extent of the disk is consistent among the three observations. The MIRI observations have a brighter peak measured closer into the central star and a flatter slope in the intermediate disk regions than either MIPS or SOFIA. Bright flux from the inner disk will be pushed outward via PSF convolution, and the discrepancies in the radial profiles are consistent with the differences in the FWHM (the MIPS PSF FWHM at 24 $\mu$m is 6.″25, the SOFIA PSF FWHM is 3″, while the MIRI F2550W PSF FWHM is 0.″803).

[4] The inner edge is chosen conservatively to begin at a separation of two FWHMs of the MIRI 25.5 $\mu$m PSF to avoid the saturated core.

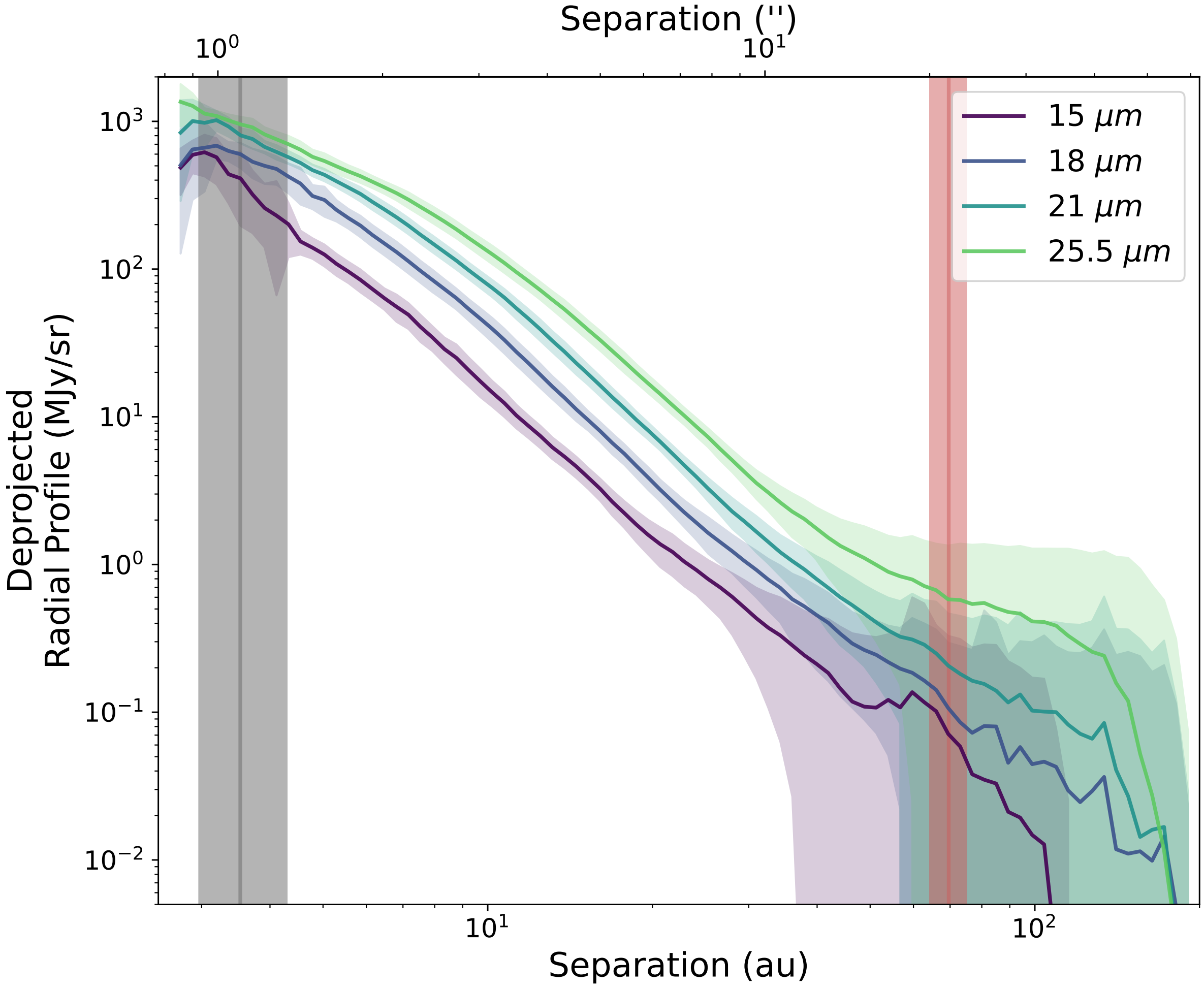

**Figure 5.** Radial profiles of the deprojected images of $\epsilon$ Eri for the four wavelengths. The vertical shaded regions represent the location of the region potentially cleared by the planet candidate $\epsilon$ Eri b (gray), and the planetesimal belt observed with ALMA (red). While a small enhancement of the disk surface brightness is possibly discernible at the location of the cold outer belt at 15 $\mu$m, this is not seen at longer wavelengths.

### *4.3. Outer Disk*

The disk is only marginally detected above the thermal background at the known location of the outer ring. ALMA observations of the ring (M. Booth et al. 2023) show it to be well confined, similar to the outer ring around Fomalhaut (M. A. MacGregor et al. 2017). In both cases, the narrowness of the ring is hypothesized to result from confinement via dynamical interaction with as yet undetected planets (see G. M. Kennedy 2020, and references therein). In the case of $\epsilon$ Eri, clumps detected with ALMA are inferred to result from a planet (near ∼40 au) migrating outwards, trapping particles into sweeping mean-motion resonances (M. Booth et al. 2023). However, we do not detect any disk substructures at that separation that would indicate a planet, potentially arguing against the migration scenario.

While the Fomalhaut and $\epsilon$ Eri outer planetesimal belts appear very similar at millimeter wavelengths, there is a clear departure extending from the optical into the mid-IR. The Fomalhaut disk is clearly visible with MIRI at 25.5 $\mu$m (A. Gáspár et al. 2023) with a surface-brightness enhancement at (and just interior to) the location of the outer belt. It is expected that interaction with an unseen planet is able to stir the outer ring (e.g., A. J. Mustill & M. C. Wyatt 2009), producing collisions that generate a large population of small dust grains (e.g., P. Kalas et al. 2005; A. C. Quillen 2006). In contrast, no surface-brightness enhancement at the $\epsilon$ Eri outer belt location is seen in these data or at shorter wavelengths (with HST at optical wavelengths and with near-IR polarimetry with SPHERE; S. G. Wolff et al. 2023; P. M. S. Krishnanth et al. 2024; C. Tschudi et al. 2024). This paucity of small dust grains resulting from a collisional cascade points to the diversity of these systems.

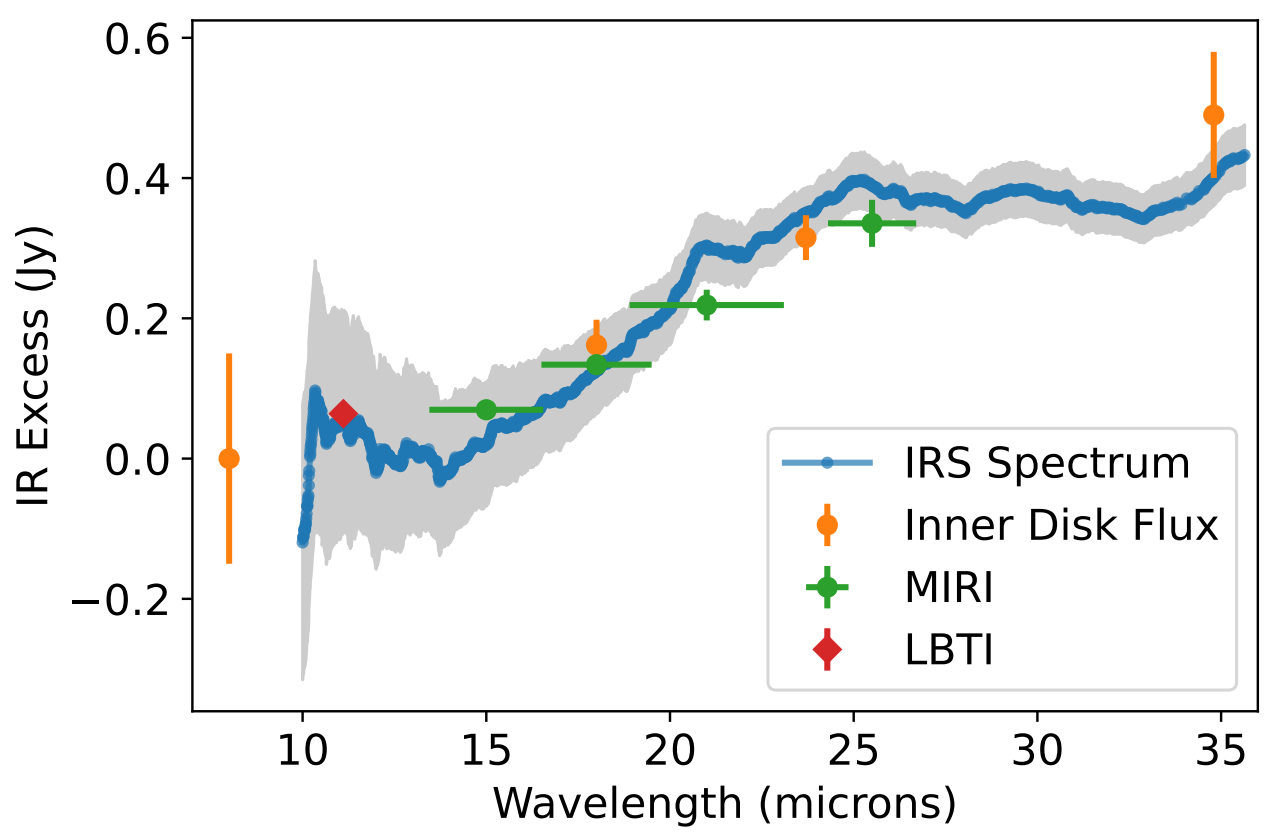

**Figure 6.** Derived disk photometry from the four MIRI imager observations are compared to the photosphere-corrected Spitzer/IRS spectrum (K. Y. L. Su et al. 2017) and flux estimates available in the literature corrected to only include inner disk flux as described in S. G. Wolff et al. (2023). Values agree to within $1\sigma$. The red diamond is derived from HOSTS survey nulling interferometry (S. Ertel et al. 2020), corrected approximately for the interferometer response. It is insensitive to any emission outside about $1''$.

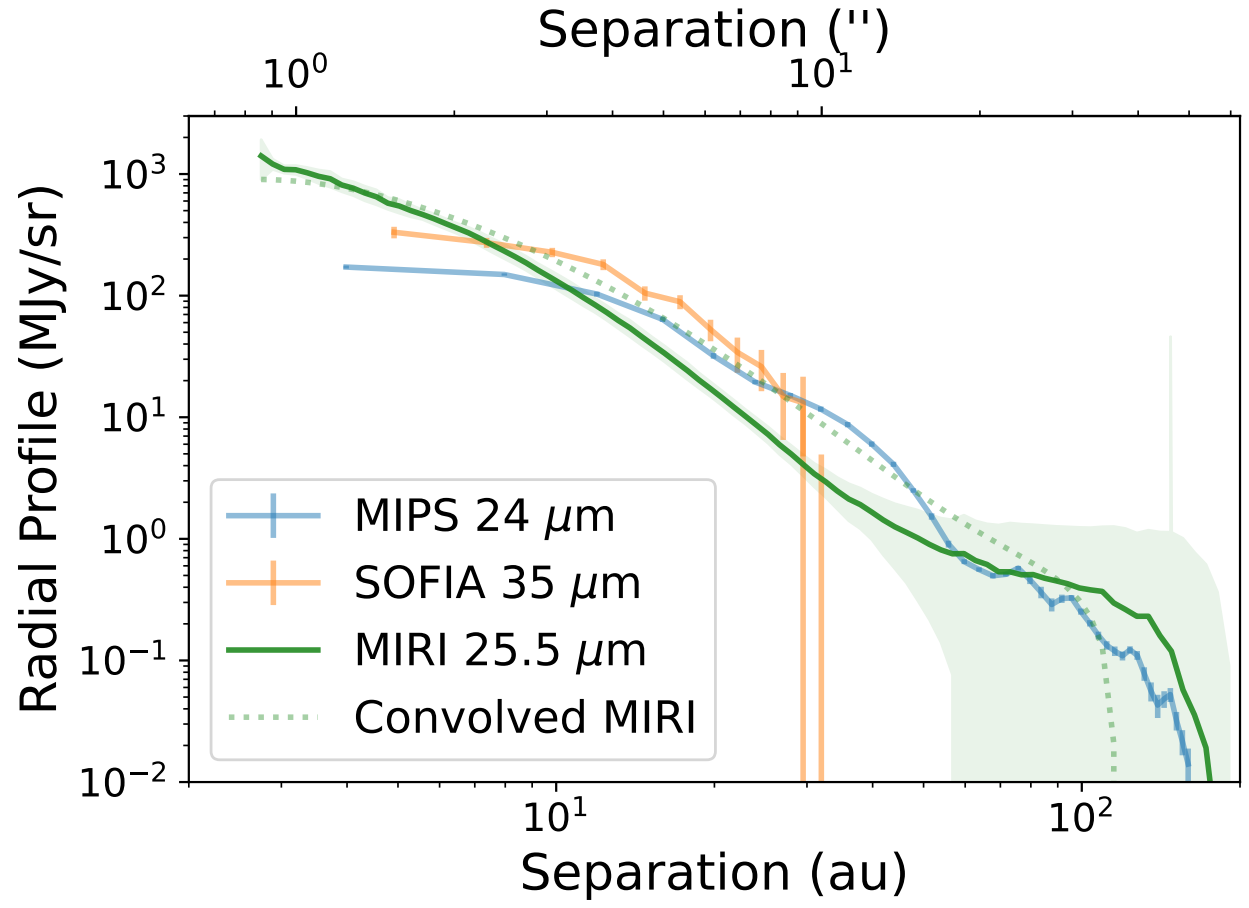

**Figure 7.** The radial profile for the MIRI 25.5 $\mu$m observations are compared to the MIPS and SOFIA radial profiles. The outer extent of the disk is consistent. While the MIRI profile shows a brighter peak closer into the central star and is fainter in the outer regions of the disk, this can be explained by PSF convolution. The dotted line shows the MIRI profile smoothed by a 1D Gaussian filter consistent with the MIPS 24 $\mu$m PSF FWHM.

There are two possible explanations for this nondetection. (1) The $\epsilon$ Eri outer belt may be undergoing a period of quiescence. Collisions may be rare with a timescale longer than the dominant drag timescale in the disk and most, if not all, small grains produced in collisions may have already been dragged into the central star. (2) This is a sensitivity issue. The predicted radiative equilibrium temperature for the outer disk centered at 70 au is only 25 K (compare to ∼48 K for Fomalhaut; M. A. MacGregor et al. 2017). Even the longest-wavelength MIRI observations are not strongly sensitive to such cold dust.

Counterintuitively, the tentative enhancement of the observed surface brightness at the location of the outer belt is strongest at our shortest wavelength, 15 $\mu$m. This is likely due to the lower thermal background at the shorter wavelengths. Deeper observations optimized for this outer region (rather than to minimize saturation near the inner disk) with JWST could confirm the detection and mid-IR structure of the outer belt, providing an estimate of the dust production rate.

### 4.4. Modeling of the Disk Inner Edge

In this section, we investigate what limits can be placed on the inner disk structure using these data and how this relates to the known planet, $\epsilon$ Eri b. Using the method described in D. Mawet et al. (2019) to define the chaotic zone resulting from resonance overlap in the low-planet-eccentricity case ($e < 0.3$; see Table 1; S. Morrison & R. Malhotra 2015), J. Llop-Sayson et al. (2021) predict a cleared, chaotic zone for the $\epsilon$ Eri b planet from 2.97 to 4.29 au. Applying the same method to the updated planet mass from W. Thompson et al. (2025), we predict a region cleared of dust extending from 2.91 to 4.29 au. The disk is detected down to our saturation limit at ∼3 au, inside of the predicted semimajor axis for the exoplanet $\epsilon$ Eri b (3 au corresponds to a separation of 0.″9 compared to the PSF FWHM at 25.5 $\mu$m for MIRI of 0.″8). However, some of the flux in this region may result from warm inner disk signal. $\epsilon$ Eri b has never been directly imaged, despite deep searches (SPHERE, NIRCam; C. Tschudi et al. 2024; J. Llop-Sayson et al. 2025). Furthermore, the inclination of the exoplanetary orbit may be offset from that of the disk (W. Thompson et al. 2025), and we would expect to see signatures of the planet imprinted in the disk's surface density (e.g., A. A. Sefilian et al. 2025).

Evidence for some inner disk structure in these MIRI observations can be found in the PSF residuals extending outward along "spokes" in each image. These are observed at all wavelengths, but are most prominent at 18 and 21 $\mu$m. If these PSF residuals were the result of an undersubtraction of the contributions from the central star during the reference differential imaging steps described in Section 3, we would expect (1) the speckle pattern to be better resolved within the diffraction spikes as is observed for point sources, and (2) the MIRI integrated photometry shown in Figure 6 would be overpredicted compared to photosphere-corrected IR values available in the literature. The integrated flux from our 15 $\mu$m image is inconclusive in this regard.

However, the HOSTS program using the Large Binocular Telescope Interferometer (LBTI; S. Ertel et al. 2020) found a compact source of flux at $\lambda = 11.11$ $\mu$m, $\Delta\lambda = 2.6$ $\mu$m that, in the largest field analyzed (∼1″ diameter), is ∼0.46% ± 0.09% of the photospheric flux from $\epsilon$ Eri at that wavelength. The LBTI response pattern on an extended source will detect only roughly one half of the full flux, so there appears to be a signal close to the star of ∼0.9% of the photospheric flux, or ∼64 ± 13 mJy, where the error is only the LBTI contribution. This error is a measure of the reality of the detection, but larger errors must be assumed in the actual flux estimate (e.g., due to the uncertain correction for the interferometer response pattern). The interferometer is not sensitive to emission from larger areas, so this value is a lower limit.

As a test of the inner disk structure, we perform radiative transfer modeling within an MCMC framework using the `emcee` package (D. Foreman-Mackey et al. 2018). We invoke a disk model with two rings of material: an inner ring ($R_1$) required to exist within the orbit of $\epsilon$ Eri b (with free parameters $R_{1,\mathrm{inner}} < R_{1,\mathrm{outer}} < 3$ au described the ring edge locations), and an outer disk ($R_2$) extending out to a fixed value of 130 au. Additionally, we require $R_{1,\mathrm{inner}}$ to be greater than the adopted stellar radius ($R_*$) of 0.7 $R_\odot$ (A. D. Rains et al. 2020). Rather than including the inner edge of the outer disk as a free parameter, we specify a gap width ($R_{2,\mathrm{inner}} = R_{1,\mathrm{outer}} +$ gap width) near the predicted location of $\epsilon$ Eri b. The dust mass in each ring was left as a free parameter, with the log of the inner disk mass confined to be $-22 \leqslant \log(M_1/M_\odot) \leqslant -3$ and the outer disk mass confined to be $-14 \leqslant \log(M_2/M_\odot) \leqslant -3$. The surface density of each disk is represented simply as $\Sigma(r) \propto r^p$, with the surface density exponent, $p$, left as a free parameter. The surface density exponent of the disk was kept constant across the two rings. Perpendicular to the disk midplane, we assume a relatively flat disk defined by a Gaussian vertical density profile with a scale height of 1 au at a reference radius of 100 au. While this density structure is less complex than other debris disk models, fits to the unresolved components under consideration are less sensitive to changes in the surface density distributions.

This modeling effort is designed to study the distribution of the innermost dust located within the saturated regions of our observations, and is not meant to provide a rigorous exploration of dust properties. We assume a dust population similar to that of N. P. Ballering et al. (2016) with a mix of astronomical silicate grains with optical constants adopted from B. T. Draine & H. M. Lee (1984) and amorphous olivine from A. Li & J. M. Greenberg (1997). We use a minimum grain size of 2 $\mu$m adopted from S. G. Wolff et al. (2023) and a maximum grain size of 1 mm with a size distribution represented by a power law with an exponent of 3.65 undergoing Mie scattering. This simplistic assumption of grain properties could bias the opacity, and therefore the surface brightness, of the model images. However, by allowing the dust mass and surface density exponent to vary, we are able to marginalize over these effects. The best-fit values derived for the dust masses and surface density exponent may not be unique and are not given significant weight in our analysis.

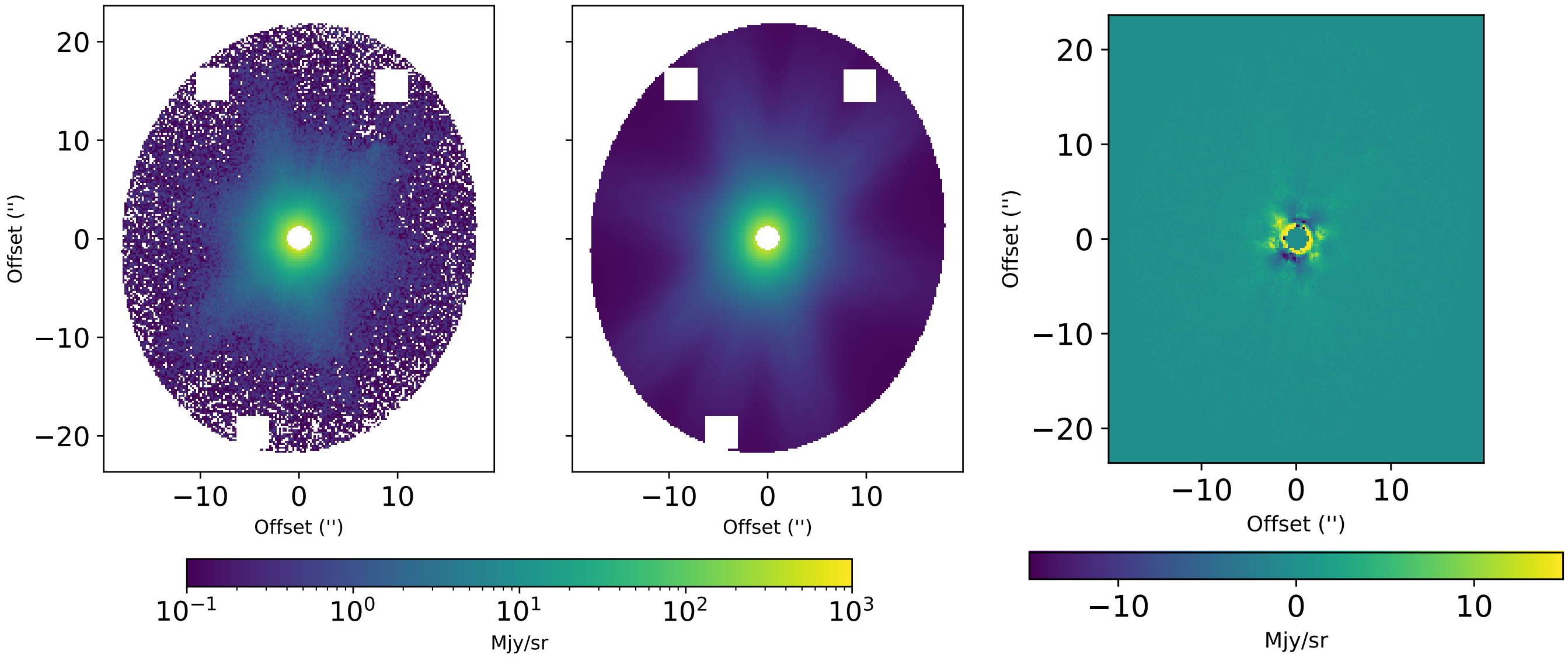

**Figure 8.** Radiative transfer modeling results for the extended disk emission and PSF residual spokes resulting from unresolved warm dust emission. The MCFOST model for the best-fit parameters (center) given in Table 3 is compared with the MIRI 18 $\mu$m observations (left). Observation – model residuals are shown on the right. Despite some core residuals resulting from an imperfect reference PSF, the model does an excellent job of fitting both the extended disk emission and the residual spokes.

For a given set of model parameters, synthetic disk images were generated at 18 $\mu$m using the radiative transfer code MCFOST (C. Pinte et al. 2006, 2022). Each image was then duplicated, rotated to match the PA of the two rolls, convolved with a high-fidelity MIRI PSF at the appropriate wavelength, rotated to north up, and median-combined. We opted to model only the F1800W image in these modeling efforts; the "spokes" are most prominent at 18 and 21 $\mu$m, and the 18 $\mu$m image provides a slightly smaller PSF FWHM. We found that including additional wavelengths in the fit did not improve the parameter constraints or convergence time to sufficiently justify the increase in the computation time per model.

The central point source is masked in the resultant model image prior to PSF convolution. The $\epsilon$ Eri disk is very faint, and the model images would otherwise be dominated by stellar PSF contributions. This approach is equivalent to assuming that a perfect stellar PSF subtraction was performed for each of the MIRI observations and is valid assuming the uncertainties from the PSF subtraction are included correctly when computing the goodness-of-fit metric within the MCMC (in this case, the $\chi^2$ value used to compute the likelihood of each model given the observations). We also mask all saturated pixels, three background point sources within the FOV, and all pixels outside a radius of 70 au where the background dominates.

We recall that the best fit to the disk contours requires a small eccentricity, corresponding to a small offset between the central star and the central point of the disk. To account for this, we include additional parameters for the $x$- and $y$-offsets in the MCMC fit. The image shifting is performed after the central star is removed to avoid interpolation artifacts from this strong signal, and before the PSF convolution step.

**Table 3**
MCFOST Disk Parameters

| Disk Parameter (units) | Priors | MCMC Results |
|---|---|---|
| $R_{1,\mathrm{inner}}$ (au) | $[R_*; R_{1,\mathrm{outer}}]$ | $1.7^{+0.2}_{-0.3}$ |
| $R_{1,\mathrm{outer}}$ (au) | $[R_{1,\mathrm{inner}}; 3]$ | $2.1^{+0.5}_{-0.3}$ |
| Gap Width (au) | [0.0; 8.0] | $0.8^{+0.4}_{-0.3}$ |
| $\log(M_1/M_\odot)$ | [−22.0; −3.0] | $-19.2^{+0.3}_{-0.3}$ |
| $\log(M_2/M_\odot)$ | [−14.0; −3.0] | $-6.7^{+0.03}_{-0.01}$ |
| Surface density exponent, $p$ | [−3.0; 3.0] | $1.06^{+0.02}_{-0.01}$ |
| $x$-offset (pixels) | [−3.0; 3.0] | $-1.43^{+0.02}_{-0.02}$ |
| $y$-offset (pixels) | [−3.0; 3.0] | $1.58^{+0.03}_{-0.03}$ |

For each set of parameters, the final median-combined 18 $\mu$m model image is subtracted from the observations. The square of these residuals is divided by the square of the uncertainties to compute the sum of the $\chi^2$ values for each unmasked pixel. The uncertainties are assumed to be dominated by Poisson noise and are computed as the square root of the observed 18 $\mu$m image. We compute the log likelihood ($P$) assuming a Gaussian normal distribution with $\log P \propto -\frac{1}{2}\chi^2$. The eight free parameters are fit via an MCMC using 35 walkers and 7000 iterations. We use the affine-invariant STRETCH MOVE sampler from J. Goodman & J. Weare (2010) and confirm convergence following their recommendation using the integrated autocorrelation time ($\tau_x$). We compute an effective sample size $N_{\mathrm{samples}}/(2\tau_x) \sim 200$ for all parameters.

The best-fit parameters are given in Table 3, the best best-fit 18 $\mu$m image and the associated observation—model residuals are shown in Figure 8, and the corner plot is shown in Figure 9. The models are able to produce the observations very well.

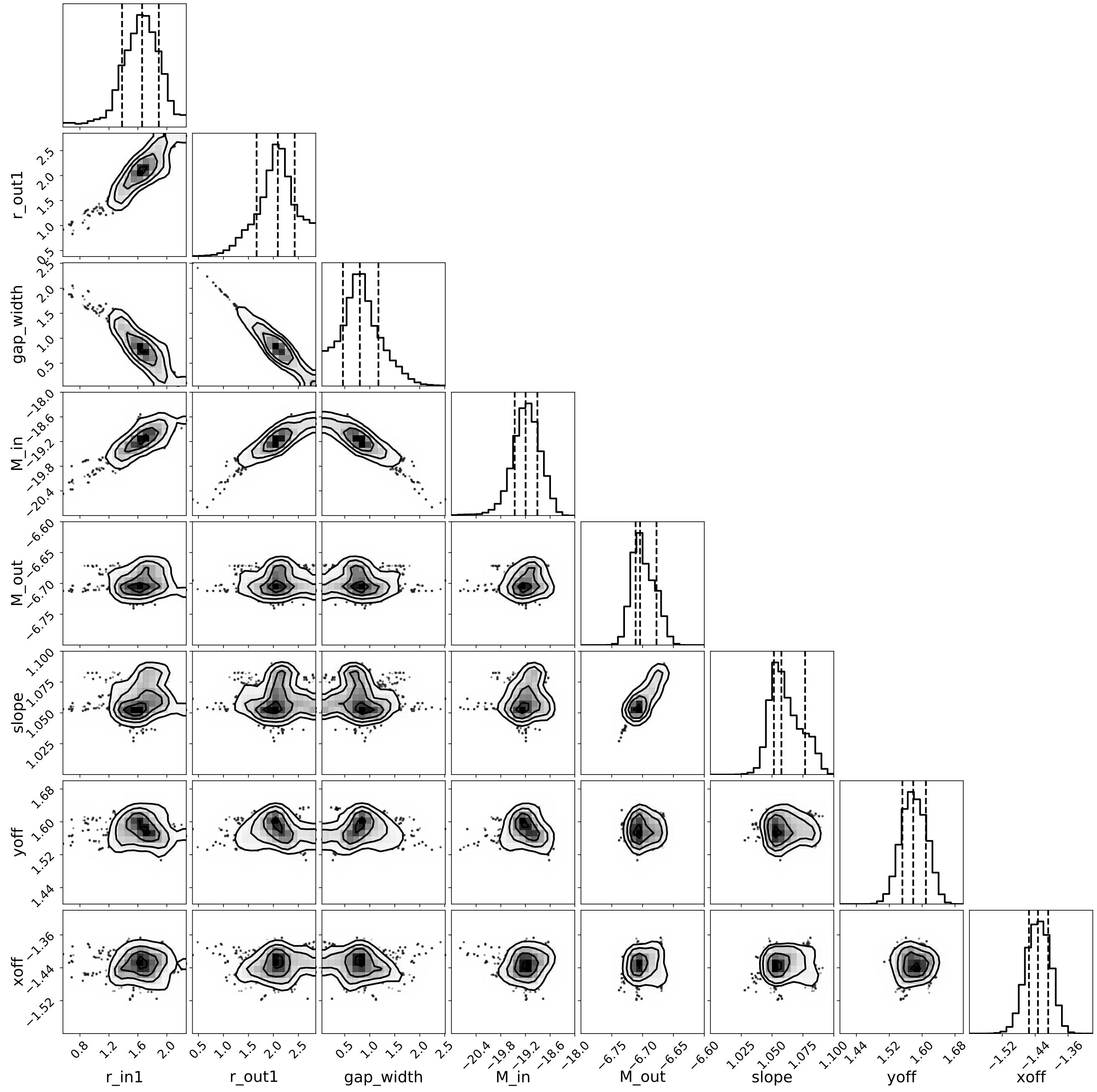

**Figure 9.** Posterior distributions from the MCMC fit to the 18 $\mu$m image using a two-disk-component MCFOST radiative transfer model. The 16%, 50%, and 84% quantiles are shown with dashed lines and listed in Table 3.

The PSFs used for convolution were not obtained contemporaneously, so some residuals are expected, particularly near the core. However, the PSF residual "spokes" resulting from the warm inner dust component and the surface brightness of the outer disk are well reproduced.

The best-fit disk model includes an inner disk extending from 1.7 to 2.1 au, followed by a 0.8 au gap and an outer disk beginning at 2.9 au, interior to the cleared zone predicted for the $\epsilon$ Eri b. There is a clear degeneracy between the outer radius of the inner disk ($R_{1,\mathrm{outer}}$) and the gap width, with smaller $R_{1,\mathrm{outer}}$ requiring larger gap widths. This maintains the best-fit inner radius of the outer belt, $R_{2,\mathrm{inner}} \simeq 2.9$ au.

Degeneracies also exist between the three parameters describing the inner disk: ($R_{1,\mathrm{inner}}$), $R_{1,\mathrm{outer}}$, and $\log(M_1/M_\odot)$. Together, these parameters define the total amount of flux resulting from this inner region. Dust closer to the central star receives more incident stellar radiation and is therefore brighter than the same population of dust particles would be if located farther out. As the inner ring location moves outward, the dust mass must increase to compensate.

At first glance, this inner disk architecture appears to be too close to the central star to align with the best-fit orbit for $\epsilon$ Eri b. However, we must also examine the best-fit inner disk center positions.

The best-fit pixel offsets are larger than expected based on the predicted eccentricity for the outer disk (see Section 4.1) and correspond to an offset of $\sim$0.75 au to the northeast (NE) for the warm dust responsible for the PSF residual spokes. Our simple disk model does not allow for brightness asymmetries in the inner disk. An offset of this magnitude has the effect of moving the farthest extent of the disk gap to 3.65 au, exterior to $\epsilon$ Eri b. One possibility is that there exists a population of dust trapped in resonance with the planet, causing an increase in surface brightness within the gap. This could masquerade as an offset to the inner disk as the model attempts to compensate both for the gap and then surface-brightness enhancement within the gap. A detailed radiative transfer model of asymmetries within the inner disk is outside the scope of this paper. However, we explore this concept further in Section 6.

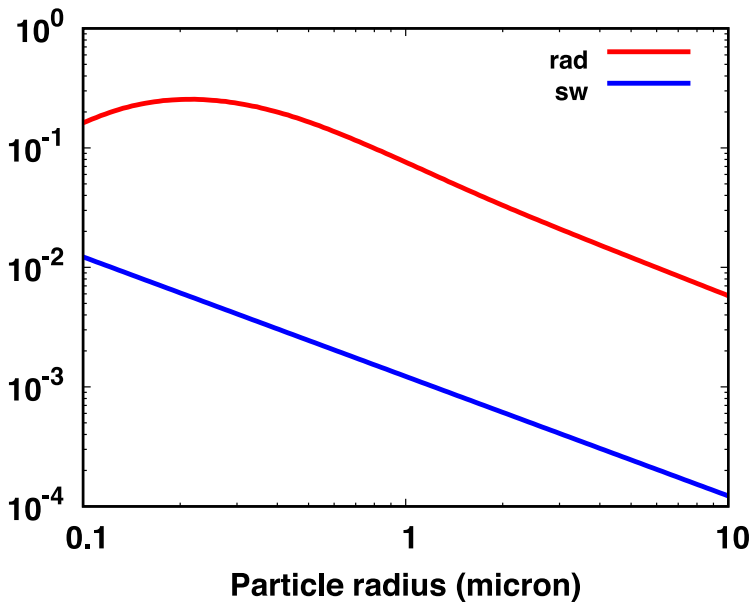

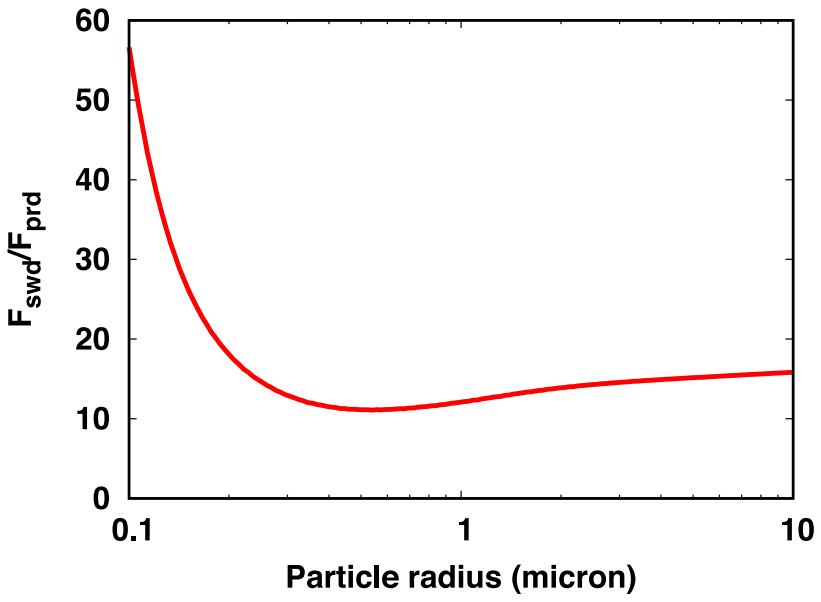

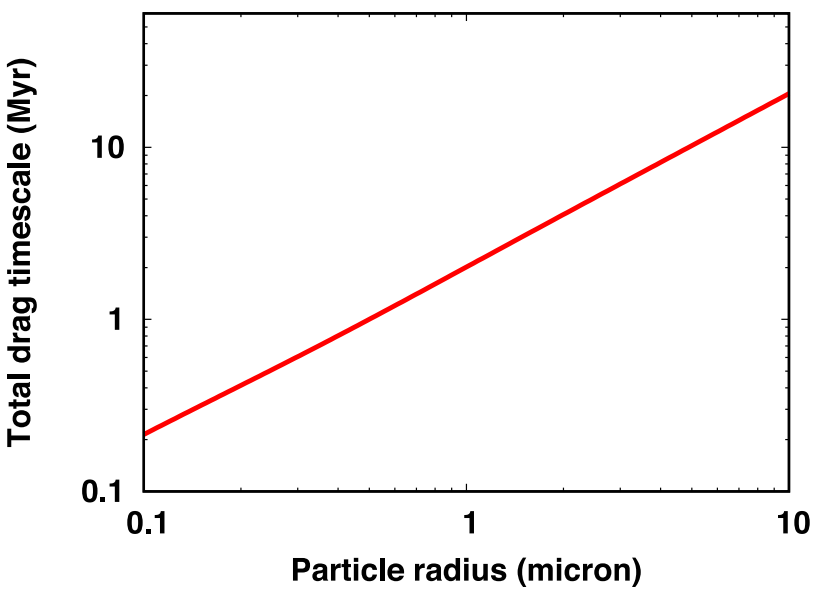

**Figure 10.** Left panel: the $\beta$ values for radiation and stellar wind (SW) calculated for the $\epsilon$ Eri system (see system variables described in the text). Middle panel: the ratio of the SW and PR drag forces in the system (note the increase in SW drag forces, due to the increase in aberration angle; see text). Right panel: the total drag timescale in the $\epsilon$ Eri system as a function of particle radius from the outer Kuiper Belt–analog ring located at 64 au, due to the combined effects of SW and PR drag.

The dust masses required to fit the inner and outer disk components are relatively low, though this value is highly dependent on our choice of the maximum particle size. The inner disk has a best-fit dust mass of $10^{-19.2}\,M_\odot$ or $2.1 \times 10^{-14}\,M_\oplus$. The modeled inner disk contributes ~23 mJy to the system SED, subject to uncertainties due to parameters such as the dust optical properties. Therefore, the measured exozodiacal contribution (see Section 4.2) is consistent with our derivations for this inner disk. The outer disk requires a larger dust mass of $10^{-6.7}\,M_\odot$ or $6.6 \times 10^{-2}\,M_\oplus$.

There is a slight degeneracy between the mass of the outer disk and the model-preferred surface density power-law slope, though both parameters are confined to a narrow range. Higher dust masses require a steeper slope weighted toward the outer disk. This has the effect of moving disk flux away from the intermediate disk regions and toward the outer belt where our observations are less sensitive. The surface-brightness distribution is best fit by a population of dust particles that increases linearly with wavelength. This is a departure from the theoretical prediction for continuous SW drag from the outer belt, as discussed in Section 5.3, and may indicate more stochastic dust production in the outer belt.

While these radiative transfer modeling results are not able to pinpoint the exact morphology of the inner disk, they confirm that a warm dust component interior to the orbit of $\epsilon$ Eri b is capable of producing the PSF residual spokes observed in these MIRI observations. Despite the prior not forcing a gap to be colocated with $\epsilon$ Eri b, our best-fit model is consistent with a gap near the planet when accounting for the best-fit offset and the limitations of the model.

## 5. Modeling the Long-term Evolution of the $\epsilon$ Eridani Debris Disk

Similar to Fomalhaut (A. Gáspár et al. 2023) and Vega (K. Y. L. Su et al. 2024), $\epsilon$ Eri's inner disk has an extended architecture. The working assumption is that it is maintained by a steady stream of smaller (micron-sized) particles being dragged inwards from the outer parent planetesimal belt(s). The dominant force responsible for this in Fomalhaut and Vega, two luminous early spectral-type stars, is Poynting–Robertson (PR) drag. For $\epsilon$ Eri, a colder, low-luminosity star, stellar wind (SW) drag is the dominant effect (see Appendix B).

To better understand the long-term evolution of the dust in the $\epsilon$ Eri disk system, we employed both dynamical models as well as separate collisional models. The dynamical models were executed to simulate the dynamical gravitational effects of $\epsilon$ Eri b as well as that of the radiative/corpuscular forces exerted by the central star, revealing the structures in the inner regions and the timescales of the perturbations. In complement to the dynamical calculations, the collisional models allow us to study the stable particle-size distributions that can exist in a system where SW drag continuously removes the smallest particles in the system.

The grain dynamics around $\epsilon$ Eri are determined by $\beta_{\rm rad}$ and $\beta_{\rm sw}$, the ratios of the radiative and corpuscular forces to the gravitational force (J. A. Burns et al. 1979):

$$\beta_{\rm rad}(a) = 0.57 Q_{\rm pr}(a) \frac{L/L_\odot}{M/M_\odot} \left(\frac{a}{\mu{\rm m}}\right)^{-1} \left(\frac{\rho}{{\rm g\ cm^{-3}}}\right)^{-1}, \quad (1)$$

and (M. J. Baines et al. 1965; J. A. Burns et al. 1979; T. Mukai & T. Yamamoto 1982; B. A. S. Gustafson 1994)

$$\beta_{\rm sw}(a) = \frac{3\dot{M} C_{\rm D} v_{\rm sw}}{32\pi G M_* \rho} \frac{1}{a}. \quad (2)$$

A detailed description of the grain dynamics, including an analysis of the individual forces, timescales, variables, and equations of motion, can be found in Appendix B of this paper. For the stellar mass-loss rate, $\dot{M}$, we adopt the value determined in K. G. Kislyakova et al. (2024), who measured the soft X-ray emission of $\epsilon$ Eri and arrived at a SW mass-loss rate of $\dot{M} = 15.6 \pm 4.4\,\dot{M}_\odot \simeq 3.1 \times 10^{-13}\,M_\odot\ {\rm yr}^{-1}$. The X-ray emission is produced when charge exchange between atoms in the wind and ISM produces an excited atom that emits a soft X-ray when it decays. B. E. Wood et al. (2002) used a different approach to estimate $\dot{M} = 30\,\dot{M}_\odot \simeq 6 \times 10^{-13}\,M_\odot\ {\rm yr}^{-1}$. Their method is based on measurement of the wings of the stellar Ly$\alpha$ emission line, which are also sensitive to the products of charge exchange between the SW and the ISM. The two methods agree within errors. We adopt the lower value from K. G. Kislyakova et al. (2024).

In the left panel of Figure 10, we plot the two $\beta$ values calculated around $\epsilon$ Eri, assuming an astronomical silicate

(B. T. Draine & H. M. Lee 1984) composition for the grains and a bulk density of 3.5 g cm$^{-3}$. The values of $\beta_{sw}$ are significantly below the values of $\beta_{rad}$ even for a star with considerable mass loss like $\epsilon$ Eri. The figure shows that particles are not removed from the system via radiation or corpuscular pressure forces, as the sum of the two $\beta$ factors remains below 0.5 (J. A. Burns et al. 1979) for this grain composition and density.

In the middle panel of the figure, we plot the ratio of the SW and radiation drag forces, for the particles we model around $\epsilon$ Eri, assuming a constant SW velocity of 400 km s$^{-1}$. As discussed in the Appendix, because the wind particles are at subrelativistic velocities, their aberration angle is large and the drag force from SW will be 10–20 times higher than from photons for all particles larger than half a micron in radius. It will be even larger for smaller particles, which will quickly spiral into the central star. That is, the SW drag timescale will be 10–20 times shorter than the PR drag timescales around $\epsilon$ Eri (see Equations (B12)–(B14)). In the right panel of Figure 10, we plot the combined drag timescale as a function of particle size in the $\epsilon$ Eri system from the inner edge of the outer Kuiper Belt–analog ring at 64 au into the central star. While the smallest sub-micron-sized particles will be dragged in within a few hundred thousand years, particles on the order of 1 $\mu$m in radius will take about 1–2 Myr to be dragged in. Therefore, the size distribution of the particles at any given distance may depend on their production rates and their drag timescales to the specific location. We expand on this topic via analytic calculations in Section 5.3. We must note that our results are approximations, assuming a steady stellar mass-loss rate, a constant (radially independent) SW velocity, and that the particle orbital velocities are much smaller than those of the charged particles in the SW. A more complete model would take these additional effects into account.

**Table 4**
DiskDyn Parameters

| DiskDyn Parameter | Value |
|---|---|
| RK4 integration time step | 0.012715 yr |
| Stellar/System Parameters | |
| Stellar mass | 0.82 $M_\odot$ |
| Stellar temperature | 5084 K |
| Stellar radius | 0.735 $R_\odot$ |
| Stellar wind mass-loss rate | 3.1 $\times$ 10$^{-13}$ $M_\odot$ yr$^{-1}$ |
| Stellar wind speed | 400 km s$^{-1}$ |
| System distance | 3.212 pc |
| System PA | −4° |
| System inclination | 33°7 |
| $\epsilon$ Eridani b Parameters | |
| Inclination | 0° and 10° |
| Semimajor axis | 3.53 au |
| Eccentricity | 0.06 |
| Mass | 1 $M_{Jup}$ |
| Grain Properties | |
| Bulk density ($\rho$) | 3.5 g cm$^{-3}$ |
| Optical composition | Astrosilicates |
| Parent Belt Disk Properties | |
| Inner radius | 64.35 au |
| Outer radius | 74.85 au |
| Disk aspect ratio ($dh/dr$) | 0.087 |
| Eccentricity distribution ($\sigma_e$) | 0.1 |
| Radius of largest body | 1 cm |
| Radius of smallest body | 0.1 $\mu$m |
| Differential size distribution slope | 3.67 |
| Disk mass | 4.28 $\times$ 10$^{-3}$ $M_\oplus$ |
| Surface density profile slope | 0 |
| Number of tracer particles | 4 million |

### *5.1. Dynamical Modeling*

To understand the spatial distribution of the dust in the inner regions of the $\epsilon$ Eri system, we model its dynamical evolution taking into account both radiative and corpuscular forces as well as the gravitational perturbations from $\epsilon$ Eri b, using our Nvidia GPU-based code DiskDyn (A. Gaspar 2025). We assume the dust to originate in the outer parent planetesimal belt (PPB), well characterized by the ALMA observations (M. Booth et al. 2017, 2023), and dragged inwards by the SW drag, meanwhile gravitationally interacting with the inner planet $\epsilon$ Eri b. We adopt the orbital parameters for $\epsilon$ Eri b from W. Thompson et al. (2025). While the existing disk observations do not provide an absolute measure of the inclination, here we assume the disk is oriented nearly coplanar with the planet.

Importantly, DiskDyn does not include particle collisions, therefore we cannot estimate the loss and production of particles. While the omission of losses is likely not important, as most larger particles will be able to travel inwards unimpeded once they leave the birth ring (see a similar calculation for Vega in K. Y. L. Su et al. 2024), we do not model the continuous production of these particles and therefore are not able to provide a completely accurate spatial density distribution. However, our model enables us to study the various timescales involved in the dust transport, as well as structures that form as a result of the dynamical perturbations from the inner planet in the system. DiskDyn also calculates the scattered light and thermal emission of the dust particles and a complete system SED, which we can compare to the observations to study (for example) the importance of not including collisional production.

The DiskDyn models were executed on the Ocelote high-performance computing (HPC) cluster at the University of Arizona on Nvidia Tesla P100 PCIe 16 GB GPUs. Due to uncertainties in the relative inclination of $\epsilon$ Eri b (W. Thompson et al. 2025), we executed two simulations, one where it was set at $\iota = 0°$ and one where it was at $\iota = 10°$. In Table 4, we summarize the variables of the model and their values in our simulations. The particle eccentricities are assigned with a normal distribution centered around zero, and the semimajor axes within values of $R_{in}/(1 - e)$ and $R_{out}/(1 - e)$. This results in a radially smoothed out distribution of particles. The true anomalies of the particles are assigned a random value between zero and $2\pi$, while inclinations are also chosen from a normal distribution, centered around zero, with a standard deviation of *dhdr*. We ran our models to 10 Myr of evolution to cover the transport of all particles smaller than 10 $\mu$m into the inner regions of the system (see right panel of Figure 10). We set DiskDyn to generate images at all observed MIRI wavelengths and at a few other wavelengths (0.6, 2.1, 4.44, 850 $\mu$m) where the system has been observed, as well as SEDs to compare to the observations. The simulations ran for approximately 30 days

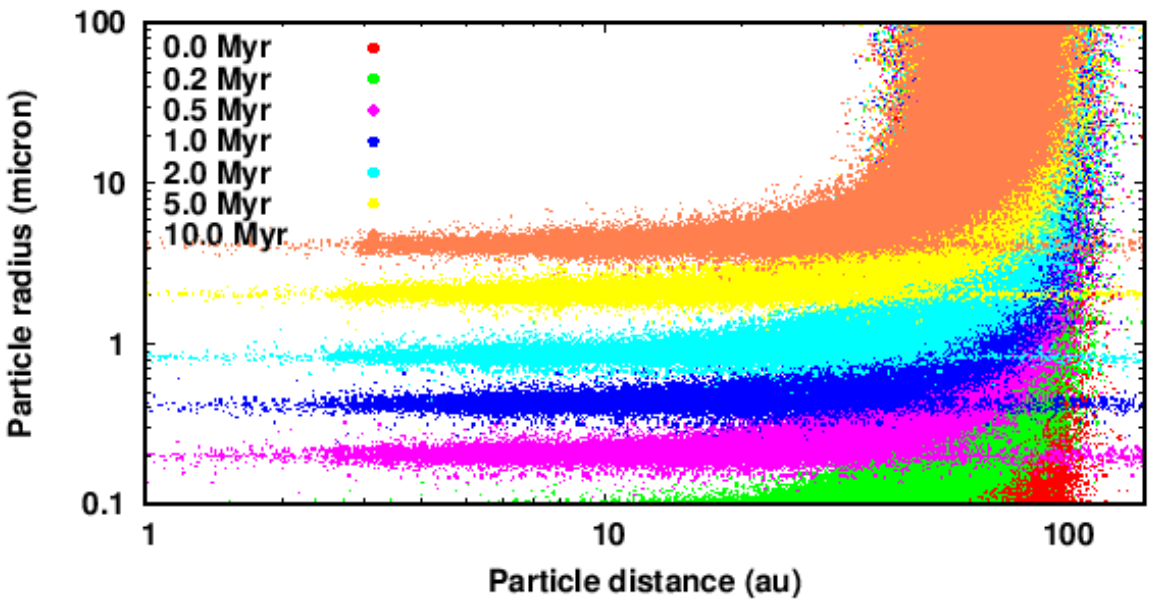

**Figure 11.** The evolution of the radial distance of the particles as a function of particle size, simulated using DiskDyn.

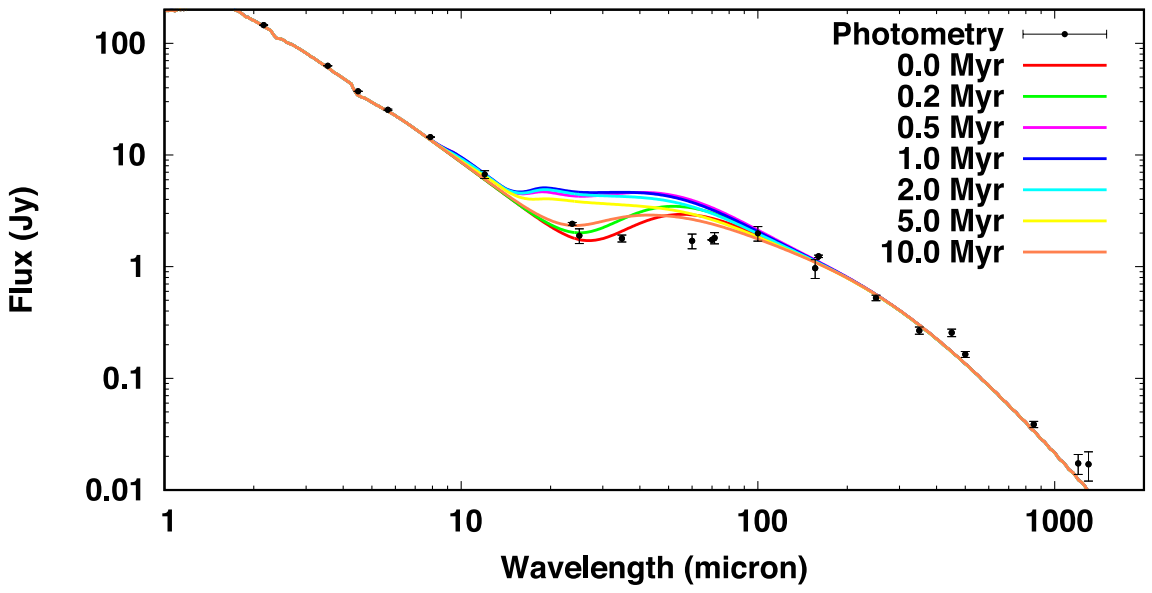

**Figure 12.** The SED of the DiskDyn models over the 10 Myr evolution of our models. We plot broadband photometry values from S. G. Wolff et al. (2023), with the inner disk flux values included (see Tables 2 and 4 in their paper).

on the HPC cluster for each of the simulations, in parallel on separate nodes.

The dynamical evolution of the *N*-body models progresses on a timescale as expected based on the analytic predictions. In Figure 11, we show the location of dust particles as a function of their radii over time. At $t = 0$ yr, the bulk of the particles are located within their birth ring, between 64 and 74 au, with particles with larger eccentricities slightly extending the ring both internally and externally (see description above). Within 0.5 Myr of evolution, the smallest particles ($a < 0.2\ \mu$m) have migrated in to the center of the system, with a radial spread based on their exact location of origin in the PPB. This migration progresses, with larger and larger particles reaching the inner regions. By 10 Myr, even 5 $\mu$m particles have migrated to the inner regions of the system. Since our model does not include the continuous collisional production of dust, at any given radial location there will naturally be a range of particle sizes that are present. The regions inwards of $\epsilon$ Eri b are cleared of particles; this is more apparent later on, when the larger particles reach this location, whose travel times are longer and therefore more under the gravitational influence of the planet. The migration results in the evolution of the SED of the system. At $t = 0$ yr, only the PPB exists, therefore only a cold blackbody emission is observed (Figure 12). As the smaller sub-micron-sized particles migrate inward, which are especially effective at emitting at near-IR wavelengths, the 2–30 $\mu$m range "lights up" at 0.5–2 Myr of the evolution timescale. With further evolution, the sub-micron-sized particles either sublimate or get flung out of the system by $\epsilon$ Eri, leaving only dust on the order of a few microns in size present after 5–10 Myr of evolution. The SED at this point starts to lose its large mid-IR ($\sim$30 $\mu$m) excess. By 10 Myr of evolution, only 5 $\mu$m particles are present in the inner system, resulting in a SED that is remarkably similar to what we observe with MIRI. In Section 5.2, we investigate the collisional production timescales of these sub-micron-sized and micron-sized particles and compare them to their drag timescales. In Figure 13, we show snapshots of the simulation at 25.5 $\mu$m for both planetary inclination scenarios, at MIRI pixel scale resolution ($0\rlap{.}''11$) and also convolved with a theoretical MIRI F2550W PSF using STPSF (M. D. Perrin et al. 2014). Dynamical structures are apparent for both models in the convolved images, with the Lagrangian quasi-stable orbital locations around $\epsilon$ Eri b lighting up.

### 5.2. Collisional Modeling

The balance between the removal of particles via SW drag and their production via collisions is critical to understanding the stable distribution of particles both in the outer and inner regions of the system. To estimate the collisional production timescales of the particles in the PPB that end up being dragged in, we use our 1D collisional model CODE-M (A. Gáspár et al. 2012a). We run a simple particle-in-a-box simulation, assuming the parameters of the PPB described in Table 4. We do increase the maximum particle radius in the distribution to 1000 km to allow the collisional evolution to properly progress. Such large bodies are unnecessary for the scattered light/thermal emission models produced in the DiskDyn dynamical models presented previously, but are necessary for the smooth quasi-steady-state collisional calculations. The collisional model was evolved to the approximate age of the system ($\sim$500 Myr) to simulate the quasi-steady state it could achieve. The initial mass scaling of the model was also increased to allow for system mass loss over the 500 Myr of evolution as well as variations from the initial distribution function. With the 1000 km maximum body radius in the system, we started our collisional models with a total debris mass of 6.3 $M_\oplus$, which is a scaling factor increase of 7.2$\times$ the particle-size distribution over that modeled in the dynamical models. While the system loses only 6.8% of its mass over the 500 Myr of evolution, the size distribution evolves from the initial distribution slope of $\eta_a = 3.67$, thereby requiring the high initial additional scaling factor for the size distribution of the evolved system to fit the observed SED. In the left panel of Figure 14, we show the integrated mass distribution of the system as a function of particle size. We chose this unit as it highlights the variations within the distribution more visibly than the standard $n(a)$ value. While the initial distribution is the "standard" steep 3.67 (A. Gáspár et al. 2012b), as one would expect from a collisional cascade, the continuous removal of particles via SW drag from the lower end of the distribution results in a much shallower distribution, with a slope of $\eta_a = 3.47$. Such a mechanism could explain the large variations in particle-size distribution slopes and their offsets from simple collisional cascade outcomes (A. M. Hughes et al. 2018). This shallower slope becomes the new "equilibrium" slope for all particles smaller than 1 mm in radius by 500 Myr of evolution, explaining the need for the higher mass scaling necessary for the collisional model. In the right panel of Figure 14, we show the SEDs calculated from the simulated size distributions, assuming an astronomical silicate composition (note the SED is calculated for only the grains present in the PPB in this plot). The SED of the system quickly loses its mid-IR hump within a few hundred thousand years, as the smallest particles leave the system in larger numbers than they are produced. By 1 Myr of evolution, most of the particles detected in mid- and far-IR emission have settled in their quasi-steady-state distribution. At 500 Myr, the submillimeter and radio emission of the model are in good agreement with the observations, while the mid- and far-IR profile is below what is observed. This is the part of the SED that the inner extended disk fills in.

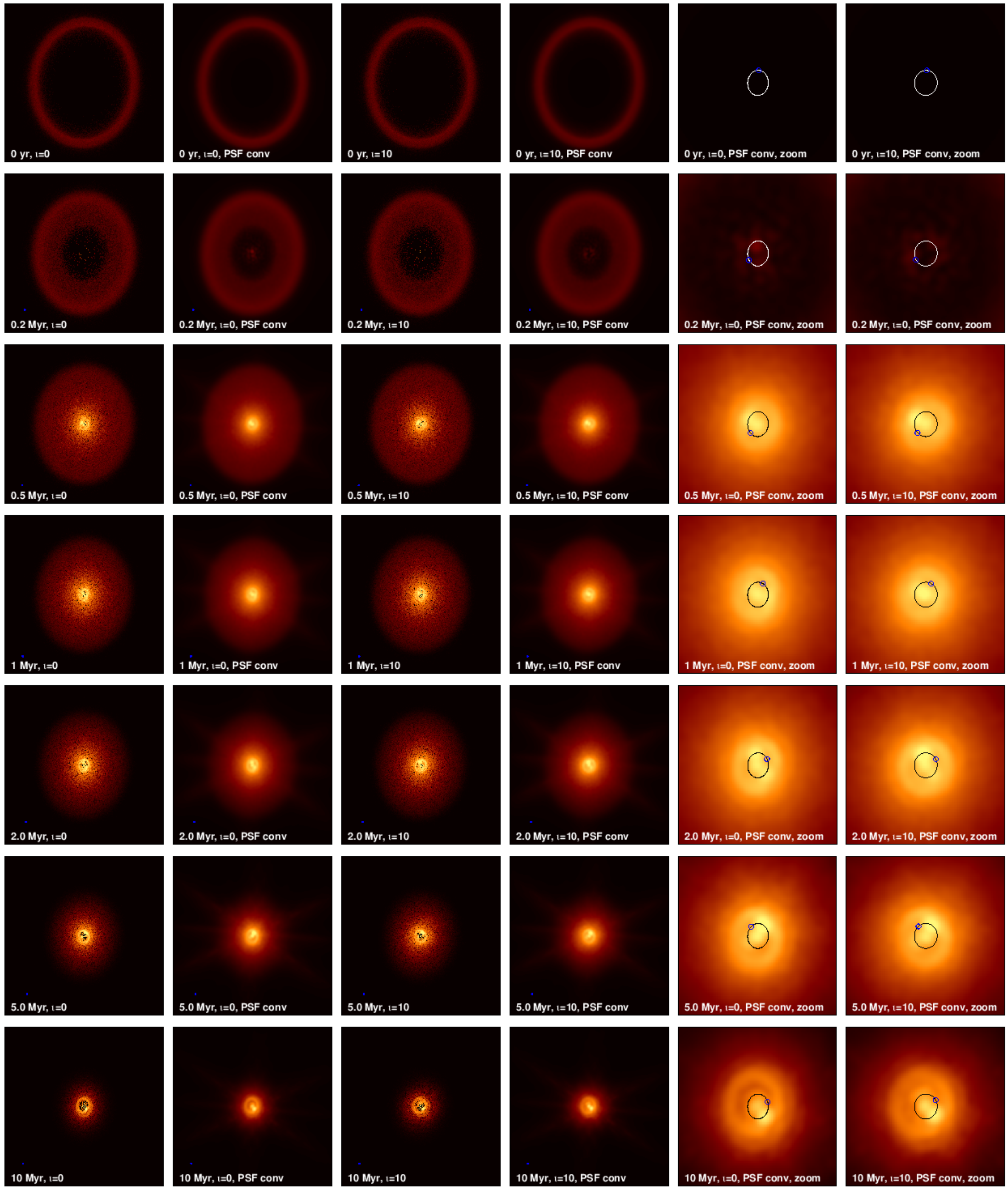

**Figure 13.** Snapshots from the DiskDyn $\epsilon$ Eri simulation at 25.5 $\mu$m (top to bottom), with the dynamic effects of $\epsilon$ Eri b and that of SW drag and PR drag included. Images are shown for both relative planet–disk inclinations modeled ($\iota = 0^\circ$ and $10^\circ$), as well as convolved with the F2550W PSF and at MIRI pixel scale resolution (0$''$11). The first four columns have a FOV of 56$''$, while the last two are zoomed in to 13$''$75. The position of $\epsilon$ Eri b is shown in the zoomed-in panels with blue circles.

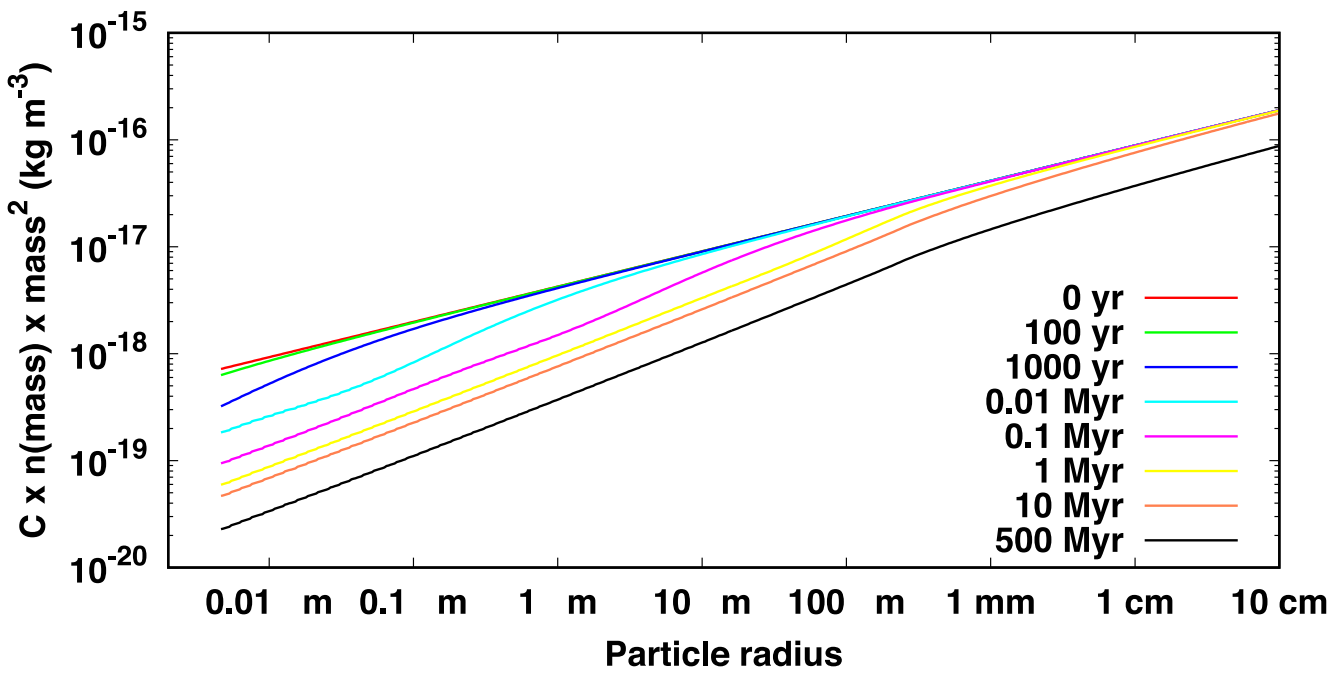

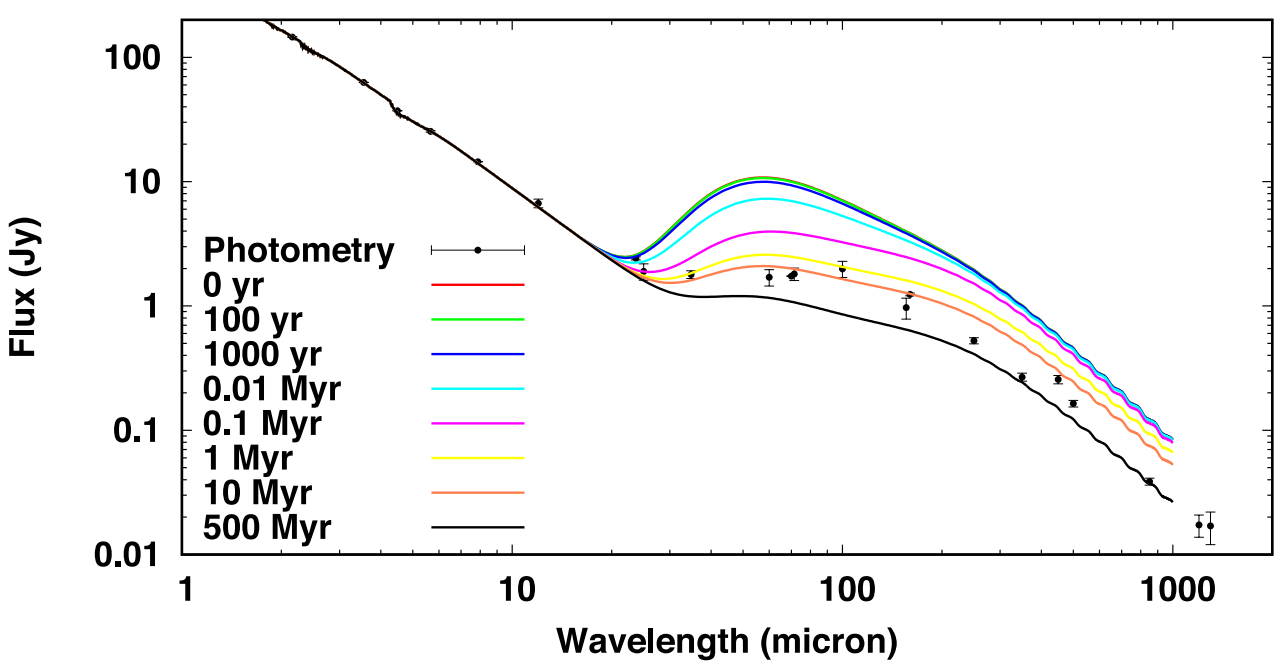

**Figure 14.** Left panel: the evolution of the particle mass distribution (multiplied by mass squared to convert to integrated mass) in the outer PPB. The absolute scaling of the distribution ($C$) depends on the binning used. Right panel: the SEDs calculated from the particle-size distributions shown in the left panel. As the smallest particles are reduced in number density, the mid-IR emission hump from the system drops, as then does the scaling of the complete excess profile as the system evolves.

In Figure 15, we show the relative change in particle number density over the collisional lifetime of the system. By 1 Myr, all particles below 20 $\mu$m in radius are in a quasi-steady-state evolution. At 500 Myr, bodies up to 100 m in radius are in collisional quasi-steady state, with a very slow evolution ($10^{-9}$ fractional losses per year). Larger particles are collisionally ground down and smaller particles removed via SW drag. In our model, the number density distribution of particles being lost after 500 Myr of evolution is fit by the following equation:

$$\frac{dn(m)}{dm} = -7.7 \times 10^{12} m^{-1.823} \text{ yr}^{-1}. \tag{3}$$

The $\eta_m = 1.823$ mass distribution slope of the loss is equivalent to a size distribution slope of $\eta_a = 3.47$. Integrating this distribution to 1 mm yields a mass loss of $6.04 \times 10^{12}$ kg yr$^{-1}$ to SW drag, which is equivalent to a solid body with a diameter of $\sim$1.5 km. As the SED of our model accurately fits the observed brightness of the disk, we anticipate the system to release approximately this amount of dust toward the inner regions (the total mass being dependent on the upper mass of the integration) every year.

### 5.3. Continuous Stellar Wind Drag and Disk Surface Density

In the previous section, we showed that SW drag and collisional cascade will reach a quasi-equilibrium in the $\epsilon$ Eri system, resulting in a continuous steady stream of particles migrating inwards from the PPB. Here, we analytically analyze the distribution of dust interior to the PPB resulting from the continuous release of dust.

The steady rate of inward flux of particles at any radial point can be described by the continuity equation

$$n(a, r)\frac{dr(a)}{dt}A = \text{constant}, \tag{4}$$

where $n(a, r)$ is the density of particle size $a$ at radius $r$, $dr/dt$ is the radial migration velocity of the particles, and $A$ is the area of the ring wall through which the particles are traversing, which can be approximated as

$$A = 2\pi r h(r) = 2\pi f r^2, \tag{5}$$

assuming that the scale height $h(r) = fr$ of the disk is linearly proportional to the radial distance, with scaling factor $f$. The migration rate of a particle, initiating its travel from an $e = 0$ orbit, can be expressed as (J. A. Burns et al. 1979)

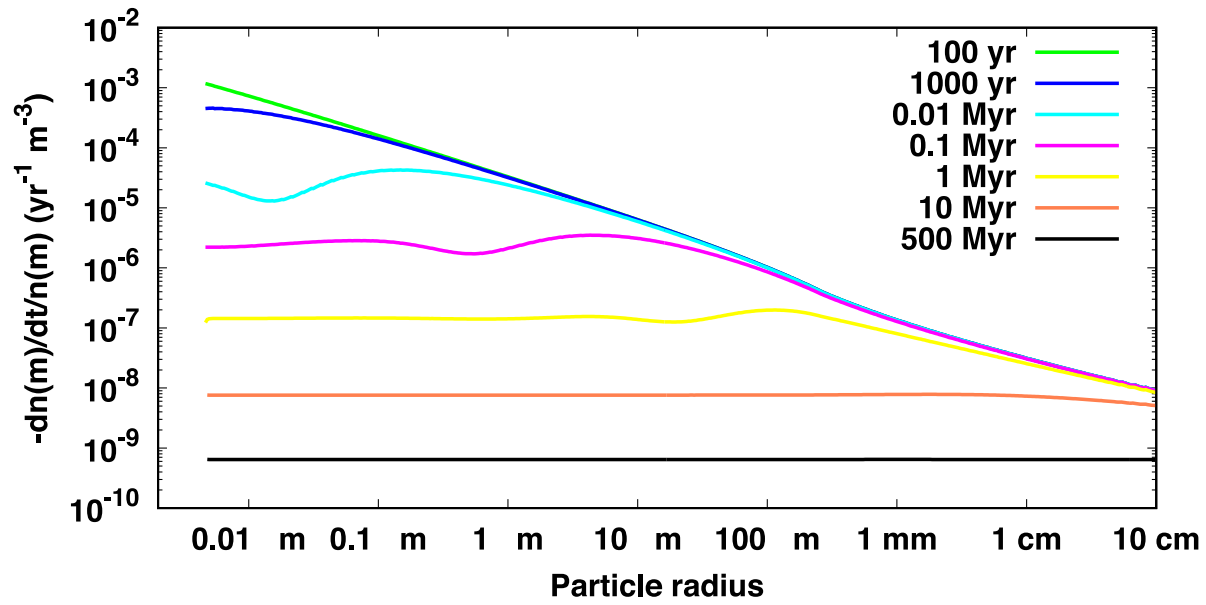

**Figure 15.** The relative change in particle number density in the parent belt over the age of the $\epsilon$ Eri system as a function of particle size. Within about 10 Myr of collisional evolution, the SW drag and collisional production reach a quasi-steady state, with the same fractional losses at all particle sizes up to 10 cm.

$$\frac{dr(a)}{dt} = -\frac{2GM\beta(a)}{c} \times \frac{1}{r}. \tag{6}$$

Substituting these into the continuity equation, we get that the number density $n(a, r)$ at any radial location will be

$$n(a, r) = n_0(a, r_0)\frac{r_0}{r}, \tag{7}$$

where $n_0(a, r_0)$ is the number density within the birth ring at radial distance $r_0$. The number density of particles increases inwards proportional to $r^{-1}$. The surface density can be expressed as

$$\sigma(a, r) = n(a, r)h(r) = fn_0(a, r_0)r_0, \tag{8}$$

i.e., the surface density will be independent of the distance from the central star (flat surface density distribution). In this case, the size distribution of the particles will be independent of location also, inheriting it from the birth ring, as long as it is in quasi-steady state.[5] However, the collisional timescale of the smaller micron-sized dust particles will be shorter than their drag timescales around $\epsilon$ Eri, therefore some deviation from this simple analytic solution is to be expected. This will be especially a concern closer to the star, where the number densities increase according to Equation (7) as well as the collisional/interaction velocities, both variables increasing the collisional probabilities. Increased collisional activity in the inner regions could lead to the steepening of the size distribution slope in the inner disk regions. Modeling the collisional and dynamical evolution of the system simultaneously is computationally expensive and beyond the scope of this paper.

[5] In the case of a flared disk, where the scale height is not linearly proportional to the disk radius, deviations from a flat surface density distribution would be expected.

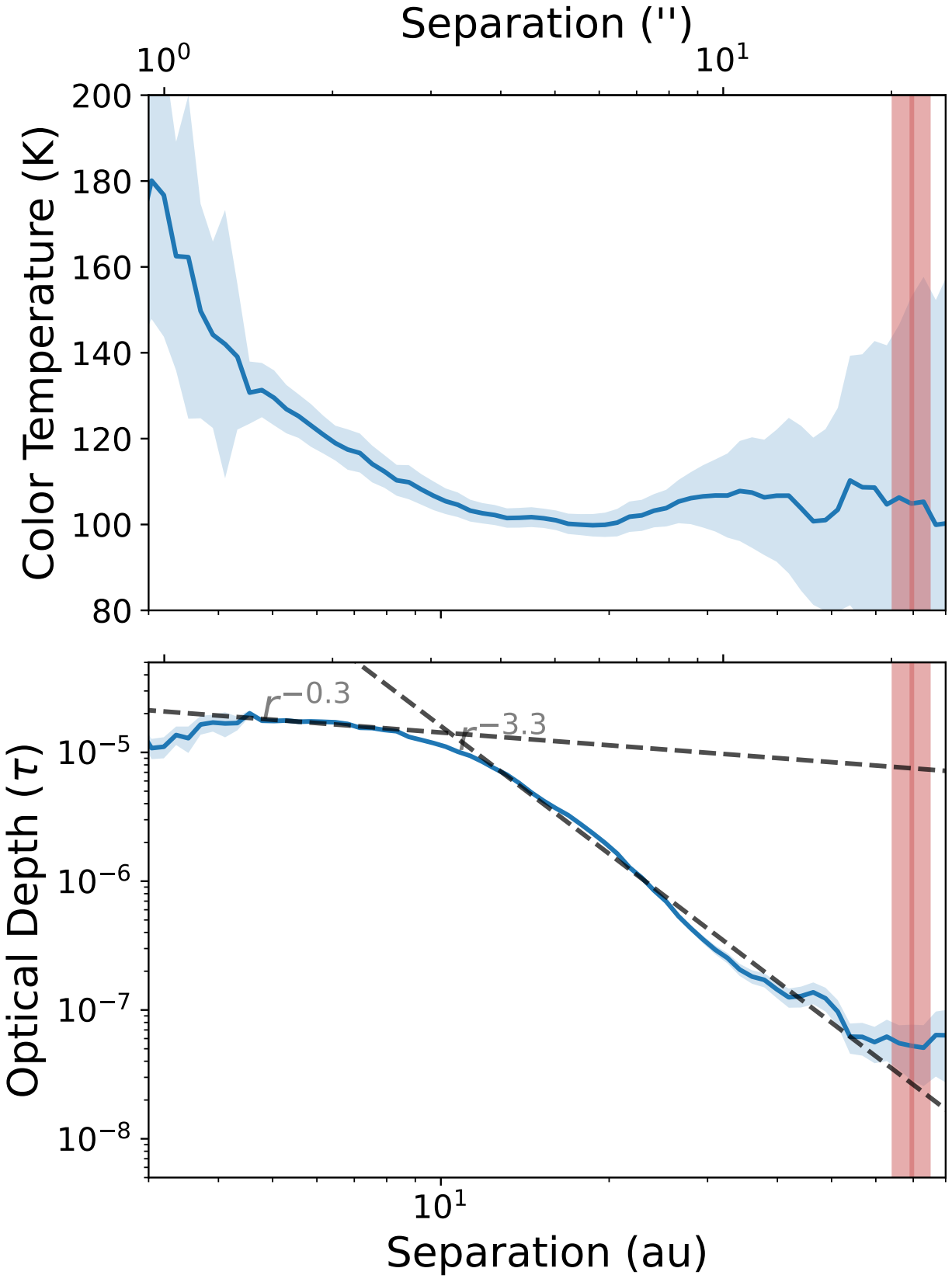

**Figure 16.** Top: the radial profile of the color temperature of the disk derived from the ratio of the F2550W and F1500W MIRI images. Bottom: the associated optical depth. For an optically thin disk, the optical depth can be approximated by the ratio of the observed flux to the product of the Planck function evaluated for the dust temperature (the color temperature above is used as a proxy for the dust temperature) and the disk solid angle.

## 6. Discussion

### 6.1. Dust Properties

The multiwavelength MIRI images allow us to place some constraints on the dust population within the disk. First, we use the ratio of the deprojected radial surface-brightness profiles between the 15 and 25.5 $\mu$m data sets (and the associated uncertainties) to estimate the color temperature of the disk. This is shown in Figure 16 (top). The temperature decreases steeply out to 20 au, where the temperature flattens out as the signal-to-noise ratio of the disk detection decreases. This is a departure from the predicted radiative equilibrium dust temperature behavior, and may indicate a change in the grain properties in the outer disk regions. However, the uncertainties in the color temperature at the location of the outer disk are large and remain consistent with a $\sim$25 K temperature within 2$\sigma$.

If we use this color temperature as a proxy for the dust temperature, the dust opacity can be estimated in the optically thin case as $F_\nu \simeq \tau_\perp B_\nu(T_{\rm dust})\Delta\Omega$, where $\tau_\perp$ is the line-of-sight optical depth, $B_\nu(T_{\rm dust})$ is the Planck function evaluated using the temperature derived above, and $\Delta\Omega$ is the solid angle (Figure 16). The opacity peaks at a value of $2 \times 10^{-5}$ at $\sim$3 au and remains fairly flat out to a radius of $\sim$10 au, where it transitions to a steeper slope with a minimum at a few $\times 10^{-8}$ in the outer disk.

The nearly flat vertical optical depth profile is expected for a disk undergoing continuous inward SW drag from a parent belt in a collisional quasi-steady state (see discussion in Section 5.3). The departure from the flat opacity profile outside of 10 au could signify a more stochastic collisional history in the outer parent belt. If the outer belt were experiencing a period of quiescence, the small dust particles with shorter drag timescales would have cleared the outer regions. The lack of a scattered-light dust detection colocated with the parent belt supports this conclusion. While a full exploration of dust particle properties is beyond the scope of this paper, we provide a qualitative comparison to the derived dust temperatures below.

The dust temperature as a function of the separation from the central star is degenerate depending on the composition and size of the dust grain. However, by combining information available in the literature, we are able to place some constraints on the dust properties. Figure 17 compares our derived dust temperature ($T_{\rm dust} = T_{\rm color}$) with the temperatures predicted for various dust species and grain sizes of 0.3 $\mu$m (top), 1 $\mu$m (middle), and 3 $\mu$m (bottom). The lack of a scattered-light detection of the disk (S. G. Wolff et al. 2023) implies a larger minimum grain size ($a_{\rm min} \gtrapprox 1\ \mu$m). Furthermore, the IRS spectrum shows no indications of a strong silicate feature (K. Y. L. Su et al. 2017).

Inside of 10 au, the measured temperature is too low to result from sub-micron-sized grains composed of astronomical silicates, amorphous carbon, or amorphous olivine. In this region, the temperature is best fit by 1 $\mu$m silicates or olivine grains or 3 $\mu$m carbonaceous dust. From 10 to 20 au, the disk dust temperature profile is flatter than for the individual dust species, and the temperature is compatible with smaller grains, though there could also be a change in the composition here. This temperature-inferred radial stratification in dust particle sizes runs counter to what was concluded based on the opacity for an ensemble of dust particles. An increasing fraction of small dust grains from 10 to 20 au is inconsistent with drag forces alone. One other possibility is that an active inner disk (interior to $\epsilon$ Eri b) could be producing a collisionally generated population of smaller dust that is being driven outward from the system via radiation pressure. Future MIRI/MRS spectroscopy would help to determine the composition and the minimum grain size present throughout this system.

### 6.2. Tension with $\epsilon$ Eridani b

Early predictions for the orbit of $\epsilon$ Eri b called for a larger eccentricity and higher inclination relative to the disk (e.g., J. Llop-Sayson et al. 2021), which would have a more disruptive effect on the dust in the inner regions. However, more recent work by W. Thompson et al. (2025) predicts that the planet is much closer to coplanar with an inclination offset $<10^\circ$ (and consistent with a disk coplanar within 1$\sigma$). For the dynamic modeling described in Section 5.1, we investigated both a coplanar and a $<10^\circ$ offset case. Qualitatively, the results look similar (Figure 13), though the inclined planet does circularize the observed disk surface brightness slightly. More intriguing is the predicted surface-brightness asymmetry within the cleared region of $\epsilon$ Eri b.

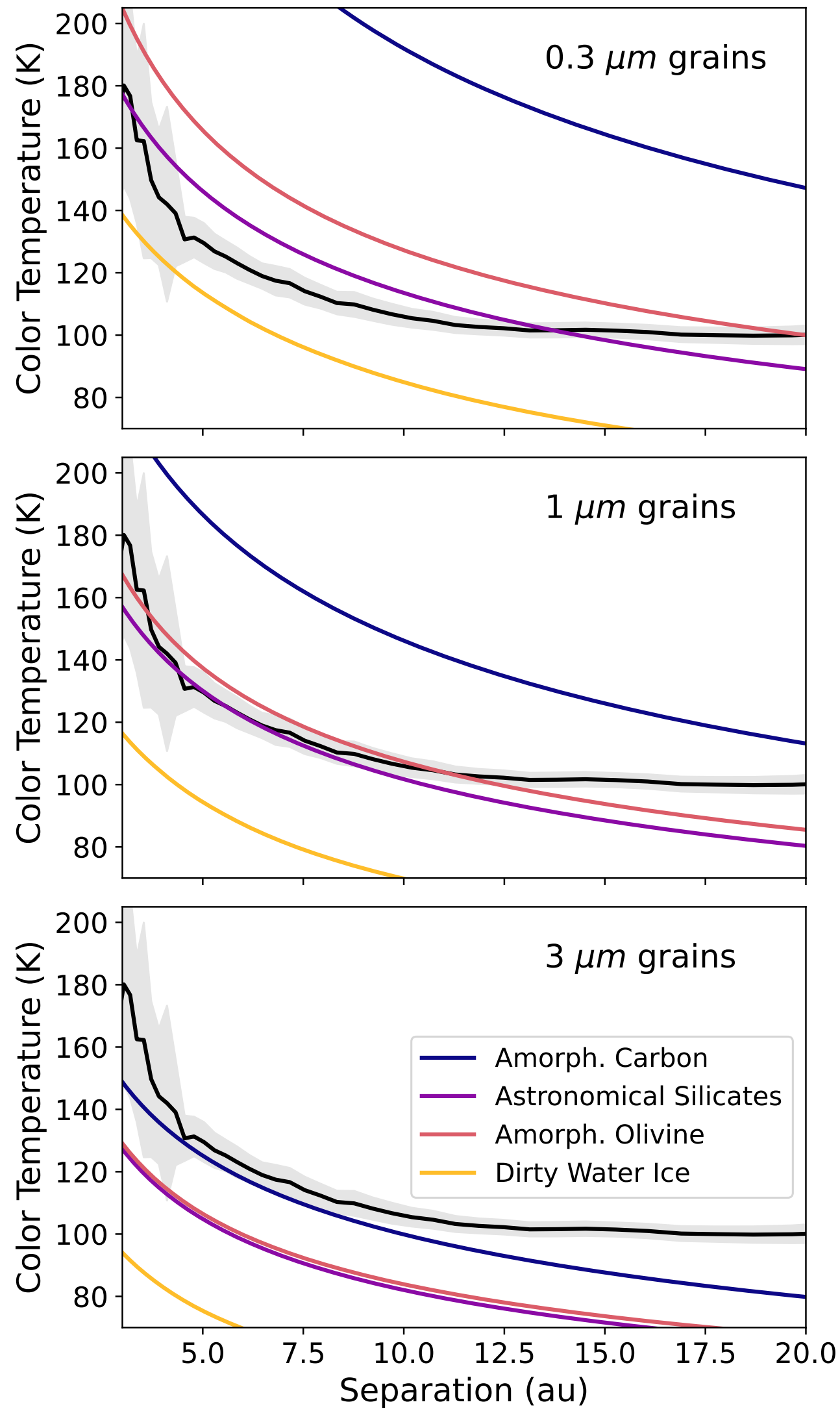

**Figure 17.** The radial profile of the disk color temperature shown in Figure 16 (black line and shaded uncertainties) is compared to the dust temperature profiles for several dust species (colors) and dust particle sizes (0.3 $\mu$m, top line; 1 $\mu$m, middle line; 3 $\mu$m, bottom). The observations are most comparable to the amorphous olivine or carbonaceous dust with grain sizes of $\sim$1 $\mu$m. Larger $\sim$3 $\mu$m amorphous carbon grains are also compatible. Water ice is too cold to dominate the dust population in the $\epsilon$ Eri disk.

Building on work by M. C. Wyatt (2003) and A. V. Krivov et al. (2007), P. Plavchan et al. (2009) provide a prescription for the probability that dust grains of a given size will be captured in mean-motion resonance with an inner planet as the dust migrates inward (see their Equation (A30)). Using the appropriate stellar and planetary parameters for the $\epsilon$ Eri system, grains of order a few microns in size are unlikely to be trapped in resonance with $\epsilon$ Eri b. However, the dynamical modeling results in Section 5.1 (see Figure 13) do show some structure in the synthetic 25.5 $\mu$m images persisting up to 10 Myr. We expect this to be a temporary pileup of dust rather than the formal 2:1 resonance capture explored analytically.

While this structure exists interior to the effective IWA of the MIRI observations (set by the saturation radius), the radiative transfer modeling results discussed in Section 4.4 do prefer an offset of the inner disk—the only allowance for asymmetric structure in the simplified disk model. During the epoch of our JWST observations, W. Thompson et al. (2025) predicted $\epsilon$ Eri b to be located to the SE. Leading and trailing resonant clumps at $\pm 60^\circ$ could explain inner disk geometry preferred by the radiative transfer modeling with an offset to the NE.

An alternative explanation for the asymmetry could be a high-eccentricity orbit for $\epsilon$ Eri b, as preferred by earlier works that did not employ Gaussian process modeling (e.g., F. Feng et al. 2023). However, a full exploration of the impact of $\epsilon$ Eri b orbital solutions is beyond the scope of this work.

Future observations are needed to pin down the orbital dynamics of $\epsilon$ Eri b and the structure of the disk interior to $\sim$3 au. There is still some tension between the semimajor axis of the planet and the best-fit gap location presented here equivalent to a $\sim 3\sigma$ disparity. Planned JWST/NIRSpec IFU Cycle 3 GO observations of the $\epsilon$ Eri system aimed at characterizing the planet (PI: Ruffio) may elucidate the source of this discrepancy. The disk is not expected to emit efficiently at these shorter wavelengths, and these observations will not place limits on the inner disk substructure.

$\epsilon$ Eri b is a potential target for future high-contrast instruments including the Nancy Grace Roman Coronagraph Instrument (expected to launch in fall 2026) and the Habitable Worlds Observatory. Of particular interest is understanding the interaction between the inner disk and the large exozodi detection in the system. At optical wavelengths, the scattered-light signal from small dust grains can interfere with the reflected-light signal of an embedded planet. The $\epsilon$ Eri system is an ideal test case for these processes.

### *6.3. Drag Forces as the Dominant Shaping Agents of Inner Debris Disks*

Decades of observations from the optical to the millimeter of Kuiper Belt analogs around nearby stars told a consistent story: PPBs were collisionally reprocessed with small dust particles spreading outward to create halos observed in scattered light. The architectures of the interior regions of the disk analogous to the asteroid belt and colocated with important ice lines have been less well characterized. Fits to IR SEDs have been unable to pinpoint the dust locations due to degeneracies with dust composition and particle sizes (e.g., P. D'Alessio et al. 2006; A. M. Hughes et al. 2018). The superb angular resolution provided by MIRI in thermal emission is uniquely capable of imaging these regions for the handful in which the thermal equivalent distances of $\sim$100 K are accessible.

The Archetypal Debris Disks GTO program (PID 1193) imaged three such systems. These $\epsilon$ Eri observations join Fomalhaut (A. Gáspár et al. 2023) and Vega (K. Y. L. Su et al. 2024) in having extended inner disk morphologies. In all cases, the observations are consistent with drag-dominated inner regions with no evidence for nascent planetesimal belts at or interior to the gas giant planet equivalent zones. For the massive A-type stars Vega and Fomalhaut, the inner regions are dominated by PR drag, while the smaller K-type star $\epsilon$ Eri is instead dominated by SW drag. Despite the small sample size, this uniform result is unexpected, and may have interesting implications for planet formation. The $\epsilon$ Eri image shows no evidence for planet-induced gaps from $\sim$3 to 50 au, though the shorter drag timescales allow for orbit crossing by $\sim$micron-sized dust.

Dust evolution in the inner disk regions also impacts the generation of exozodiacal light, the largest source of uncertainty in the future direct detection of habitable worlds (A. Roberge et al. 2012). The zodiacal cloud in our own solar system is generated from cometary and asteroidal dust, though the precise contributions remain an active area of research (e.g., M. V. Sykes & R. Greenberg 1986; J. C. Liou et al. 1995a; D. Nesvorný et al. 2010; A. Bonsor et al. 2012). The large exozodiacal population observed for $\epsilon$ Eri may be entirely composed of grains dragged in from the 70 au belt, which may result in a different composition and scattering efficiency.

Future JWST/MIRI observations to increase the sample of resolved inner disks ($\beta$ Leo and $\eta$ Crv for GTO 4538, $\eta$ Cru for GO 5650, and $\gamma$ Oph for GO 5709) will corroborate this trend and help understand the diversity of inner disk structures.

## 7. Conclusions

These $\epsilon$ Eri observations complete the sample from the MIRI GTO survey of archetypal disks. In this case, we adopted a novel direct-imaging observational approach, and showed that unocculted MIRI imager observations of the science target with a dedicated PSF reference star are able to match the contrasts provided by the various MIRI coronagraphs without the limitations in IWA and FOV from 15 to 25.5 $\mu$m. This approach is best suited to extended disk structure rather than point sources. We summarize the conclusions below.

1. We present multiwavelength direct-imaging observations of the $\epsilon$ Eri disk with JWST/MIRI. The disk extends smoothly from <3 au out to the location of the outer belt at 69 au with no signs of gaps or asymmetries indicative of planets.
2. There is no enhancement of small dust grains at the location of the outer belt, as might be expected, though these observations have limited sensitivity in the outer disk regions where the thermal background dominates.
3. PSF residuals are consistent with a warm dust population interior to 3 au. Simplified radiative transfer modeling predicts an offset dust ring from 1.7 to 2.1 au, a gap out to 2.9 au, and a smooth distribution of dust between the warm inner disk and the cold outer belt. This gap location is inconsistent with the predicted gap carved by $\epsilon$ Eri b and warrants further investigation. This JWST modeling has enabled the confirmation of a disk seen in interferometry for the first time (S. Ertel et al. 2020).
4. Dynamical modeling confirms that the $\epsilon$ Eri disk is dominated by SW drag. Small dust grains move rapidly in toward the central star, resulting in lower surface densities, and thereby explaining the lack of a detection in scattered light.
5. The balance between SW drag and the collisional cascade predicts a flat surface density distribution. This is observed out to $\sim$10 au. Exterior to this, the inferred disk temperature plateaus and the optical depth decreases rapidly, indicating a change in the dust properties and hinting at a stochastic collisional history.
6. $\epsilon$ Eri joins Vega and Fomalhaut with bright, extended inner disks consistent with a population of dust grains dragged in from an outer belt. This ubiquity was not predicted prior to the launch of JWST and has interesting implications for the physical processes that dominate in debris disk evolution. Future observations for a larger sample of debris disks are needed to place the asteroid belt in context: Are narrow belts in the inner disk regions near important ice lines uncommon? And are transport-dominated inner disks the prevailing architecture?

## Acknowledgments

Work on this paper was supported in part by grant No. 80NSSC18K0555 from NASA Goddard Space Flight Center to the University of Arizona. We thank Steve Ertel for a helpful discussion regarding the LBTI. Part of this research was carried out at the Jet Propulsion Laboratory, California Institute of Technology, under a contract with the National Aeronautics and Space Administration (grant No. 80NM0018D0004). A.A.S. is supported by the Heising-Simons Foundation through a 51 Pegasi b Fellowship.

*Facility:* JWST (MIRI).

*Software*: astropy (Astropy Collaboration et al. 2013, 2018), corner (D. Foreman-Mackey 2016), DiskDyn (A. Gaspar 2025).

## Appendix A
## Supplemental Data Reduction Material

We iteratively removed detector artifacts by first generating an initial median-combined, PSF-subtracted image. The goal is to subtract all sources of astrophysical flux to reveal underlying artifacts that can be subsequently subtracted or masked in each dither position prior to final alignment and combining. For the initial result, only the residual constant sky level was removed from the individual background-subtracted dither positions. The sky levels were calculated from regions that excluded contributions from the central PSF and were on the order of $\sim\pm1$ MJy sr$^{-1}$ in all filters. The sky constants subtracted were on the order of $\sim\pm1$ MJy sr$^{-1}$ at all wavelengths. Faint background point sources were also removed from the PSF reference images by subtracting simulated STPSF point sources at their locations. The median radial profile of the extended disk emission was then calculated from this first iteration image and subtracted from the background- and PSF-subtracted images to determine the underlying median row and column correction values. Figure 18 showcases the reduction processing steps using the 25.5 $\mu$m data as an example.

The final image, apart from including row/column artifact corrections, was generated via the same steps as the initial reduction. In A. Gáspár et al. (2023) and K. Y. L. Su et al. (2024), we showed that the MIRI long-wavelength PSF is remarkably stable, but with a significant dependence on the location on the detector. Therefore, we subtracted the PSFs taken at the same detector location from each science image. The scalings and offsets of the PSFs at each wavelength were visually determined to minimize the subtraction residuals. We applied PSF scalings of 0.805, 0.814, 0.822, and 0.840 at 15, 18, 21, and 25.5 $\mu$m, respectively. We compare this to the expected flux ratios using the JWST Exposure Time Calculator (ETC; K. M. Pontoppidan et al. 2016) to simulate both the science and PSF scenes (including only the central star with no disk component) in all four bands using the detector settings given in Table 1.[6] The ETC predicts a PSF/science scale factor of 0.803 across all bands. The "by-eye" PSF scalings may have been biased by warm dust close to the saturation region in our images, which is expected to increase in brightness with increasing wavelength. We incorporate this into the photometric uncertainties discussed in Section 4.2 and Figure 6. Custom masks were generated for each individual image at each wavelength. The eight PSF-subtracted images at each wavelength were finally median-combined, excluding signal enclosed within the custom masks. In Figure 19, we present the final eight position and rotation dithered 25.5 $\mu$m images, along with their masks, that are combined to produce the final science product of the image processing.

[6] While the ETC is not a perfect observation simulator, it correctly treats the saturation effects that dominate our images.

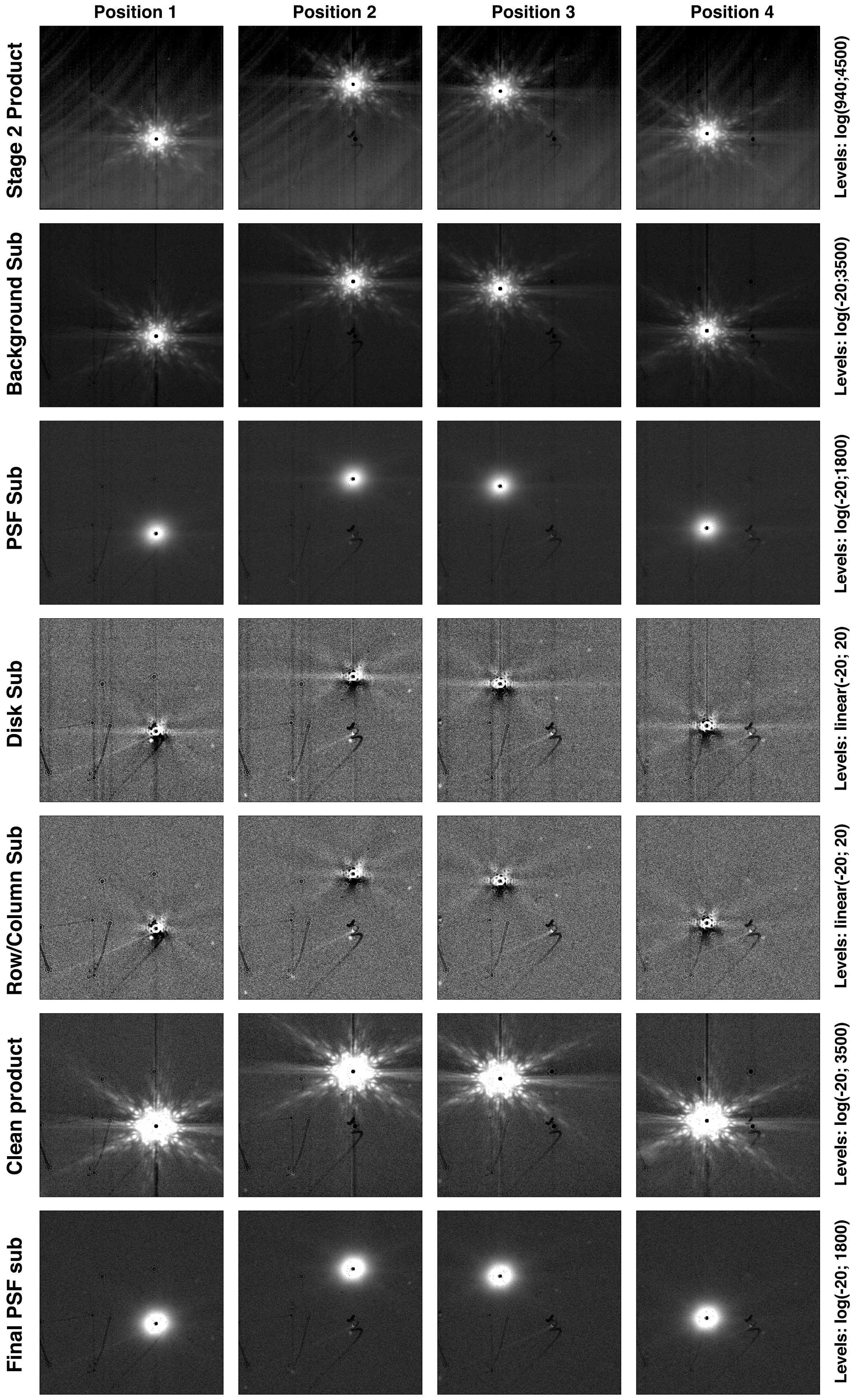

**Figure 18.** The iterative reduction sequence (top to bottom) for the $\epsilon$ Eri F2550W data set at each of the four dither positions at Rotation 1 (levels in megajanskys per steradian given on the right side of each row). The "Clean product" images contain row/column artifacts that were removed by similar artifacts present in the PSF images, as the artifacts are estimated on the PSF-subtracted images.

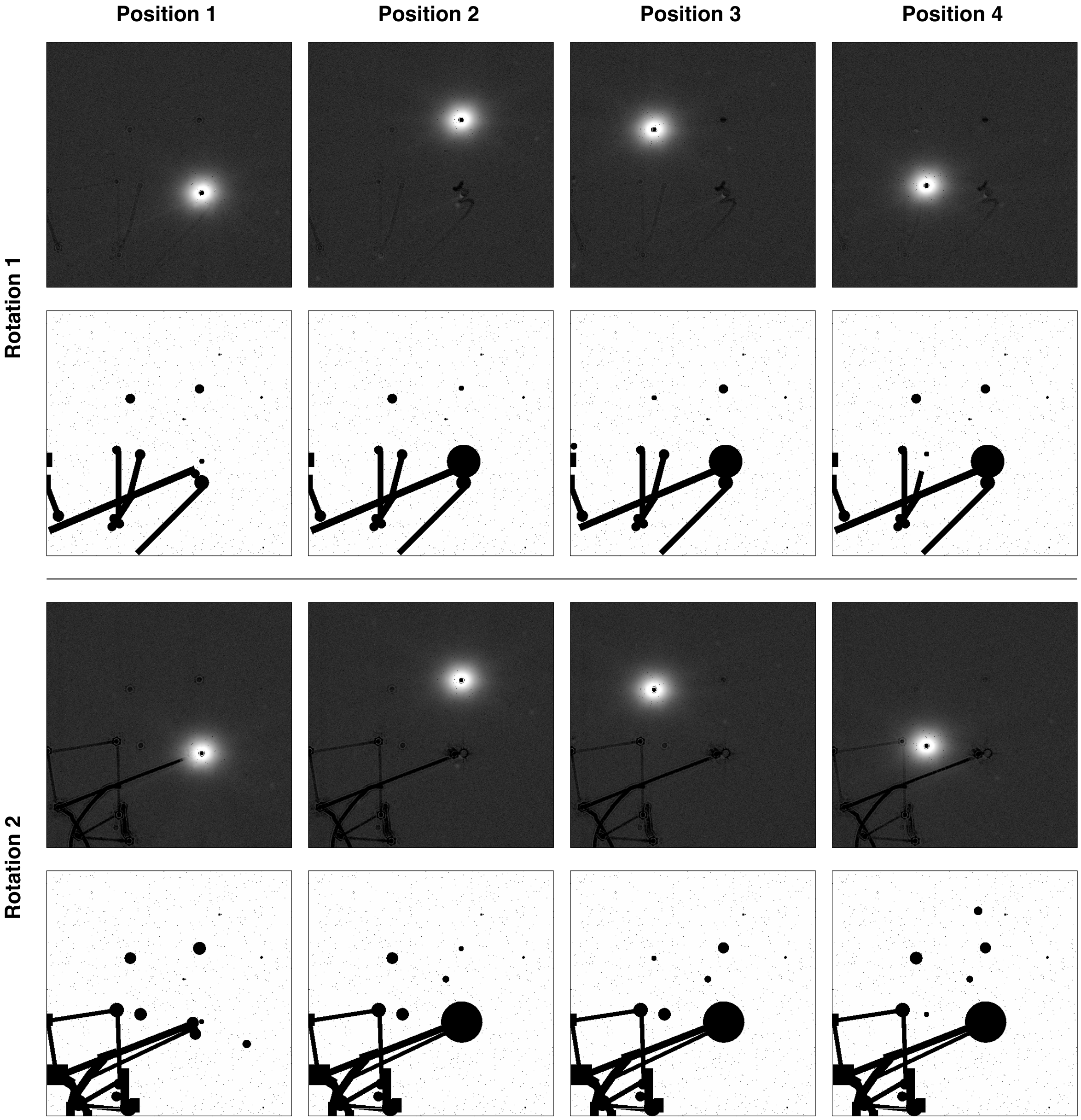

**Figure 19.** The eight final processed images at 25.5 $\mu$m, at the four extended dither positions and two rotations, shown with the masks used to produce the final combined image. Images are shown in logarithmic stretch, between −20 and 1800 MJy sr$^{-1}$.

## Appendix B
## Analysis of Dominant Forces Affecting Grain Dynamics in Debris Disks

The dynamical evolution of a dust particle with mass $m$ orbiting a star is determined by the gravitational, radiative, and corpuscular SW forces it experiences:

$$m\ddot{\boldsymbol{r}} = \boldsymbol{F}_{\rm g} + \boldsymbol{F}_{\rm rad} + \boldsymbol{F}_{\rm sw}. \tag{B1}$$

The individual forces can be expressed as (e.g., J. A. Burns et al. 1979; B. A. S. Gustafson 1994; J.-C. Liou et al. 1995b)

$$\boldsymbol{F}_{\rm g} = -{\rm G}\frac{M_{*}m}{r^2}\frac{\boldsymbol{r}}{r}, \tag{B2}$$

$$\boldsymbol{F}_{\rm rad} = {\rm G}\beta_{\rm rad}\frac{M_{*}m}{r^2}\left[\left(1 - \frac{\dot{\boldsymbol{r}}\cdot\boldsymbol{r}}{cr}\right)\frac{\boldsymbol{r}}{r} - \frac{\boldsymbol{v}}{c}\right], \tag{B3}$$

$$\boldsymbol{F}_{\rm sw} = {\rm G}\beta_{\rm sw}\frac{M_{*}m}{r^2}\left[\left(1 - \frac{\dot{\boldsymbol{r}}\cdot\boldsymbol{r}}{v_{\rm sw}r}\right)\frac{\boldsymbol{r}}{r} - \frac{\boldsymbol{v}}{v_{\rm sw}}\right], \tag{B4}$$

where $r$ is the distance of the dust particle from the central star of mass $M_{*}$, $\beta_{\rm rad}$ and $\beta_{\rm sw}$ give the ratio of the radiative and corpuscular forces to the gravitational force, $v$ is the spatial velocity of the dust particle, and $v_{\rm sw}$ is the velocity of the SW. Substituting in the expressions of each force and reorganizing, we obtain

$$\ddot{\boldsymbol{r}} = -\mathrm{G}\frac{(1-\beta_{\rm rad}-\beta_{\rm sw})M_*}{r^3}\boldsymbol{r} - \mathrm{G}\frac{M_*}{r^2} \times \left(\frac{\beta_{\rm rad}}{c}+\frac{\beta_{\rm sw}}{v_{\rm sw}}\right)\left(\frac{\dot{\boldsymbol{r}}\cdot\boldsymbol{r}}{r^2}\boldsymbol{r}+\boldsymbol{v}\right). \tag{B5}$$

The first term of this expression yields the gravitational force, reduced by the radiative and the corpuscular pressure, while the second term provides the sum of the radiative PR drag and the corpuscular SW drag. Within a Cartesian coordinate system ($\boldsymbol{x}$,$\boldsymbol{y}$,$\boldsymbol{z}$) centered on the star, this expression would be the following for the acceleration in the $x$ coordinate:

$$\ddot{x} = -\mathrm{G}\frac{(1-\beta_{\rm rad}-\beta_{\rm sw})M_*}{(x^2+y^2+z^2)^{\frac{3}{2}}}\underline{x} - \mathrm{G}\frac{M_*}{x^2+y^2+z^2}\left(\frac{\beta_{\rm rad}}{c}+\frac{\beta_{\rm sw}}{v_{\rm sw}}\right) \times \left(\frac{xv_x+yv_y+zv_z}{x^2+y^2+z^2}\underline{x}+\underline{v_x}\right), \tag{B6}$$

and by symmetry the same holds for $\ddot{y}$ and $\ddot{z}$ once the underlined are changed to the $y$- and $z$-components, respectively. The value of $\beta_{\rm rad}$ can be calculated for the particular dust grain radius $a$ and grain type as given by J. A. Burns et al. (1979):

$$\beta_{\rm rad}(a) = 0.57 Q_{\rm pr}(a)\frac{L/L_\odot}{M/M_\odot}\left(\frac{a}{\mu\mathrm{m}}\right)^{-1} \times \left(\frac{\rho}{\mathrm{g\ cm^{-3}}}\right)^{-1}, \tag{B7}$$

where $\rho$ is the bulk density of the dust material and $Q_{\rm pr}(a)$ is the radiation pressure efficiency for a given grain size, averaged over the stellar spectrum, as

$$Q_{\rm pr}(a) = \frac{\int Q_{\rm pr}(\lambda, a)B(T_*, \lambda)d\lambda}{\int B(T_*, \lambda)d\lambda}. \tag{B8}$$

The wavelength-dependent radiation pressure efficiency can be calculated from the absorption and scattering efficiencies as

$$Q_{\rm pr}(\lambda, a) = Q_{\rm abs}(\lambda, a) + [1-g(a)]Q_{\rm sca}(\lambda, a), \tag{B9}$$

where $g(a)$ is the scattering asymmetry factor, which can either be estimated using the L. G. Henyey & J. L. Greenstein (1938) approximation or by using Mie theory and calculating the scattering-angle-dependent anisotropy parameter of the dust particles and averaging it over the scattering angles. We did the latter in our simulations.

While the $\beta_{\rm rad}$ factor depends on the optical properties of the dust grains, the corpuscular equivalent only depends on the grain size $a$ and bulk density $\rho$ (M. J. Baines et al. 1965; J. A. Burns et al. 1979; T. Mukai & T. Yamamoto 1982; B. A. S. Gustafson 1994):

$$\beta_{\rm sw}(a) = \frac{3\dot{M}C_{\rm D}v_{\rm sw}}{32\pi \mathrm{G}M_*\rho}\frac{1}{a}, \tag{B10}$$

as well as properties of the central star, such as its mass-loss rate ($\dot{M}$) and the velocity of the stellar wind $v_{\rm sw}$. The $C_{\rm D}$ factor is the free molecular drag coefficient (T. Mukai & T. Yamamoto 1982; B. A. S. Gustafson 1994), for which we adopt a value of 2, assuming a "no sputtering" scenario, where protons transfer all of their momentum during interactions. According to B. A. S. Gustafson (1994), this is a valid assumption within the solar system, and we assume it to be the case for $\epsilon$ Eri.

The $\beta_{\rm sw}$ factor is a few orders of magnitude lower than the $\beta_{\rm rad}$ for all stellar spectral types and all particle sizes that are present in debris disk systems, including $\epsilon$ Eri (see Figure 10 in main text). Due to this, SW pressure cannot remove a dust particle from a circumstellar disk. The effects of drag forces are vastly different, however. While the $\beta_{\rm sw}$ factor is smaller than $\beta_{\rm rad}$, the aberration angle of the charged stellar particles, $\tan^{-1}(v/v_{\rm sw})$, is much larger than that of the photons, $\tan^{-1}(v/c)$ (note we are assuming the $v \ll v_{\rm sw}$ limit to approximate the aberration angle). This results in the drag effects from the SW force to become larger than from the radiative forces (PR drag) around stars with large mass-loss rates (such as $\epsilon$ Eri). The ratio of the SW drag force to PR drag force can be expressed as

$$\frac{F_{\rm swd}}{F_{\rm prd}} = \frac{\beta_{\rm sw}c}{\beta_{\rm rad}v_{\rm sw}}, \tag{B11}$$

which we also plot in Figure 10 in the main text for $\epsilon$ Eri, showcasing that the SW drag force is around 10–20 times higher than PR drag in this system. As a comparison, the same ratio is around 0.2–0.4 in the solar system (B. A. S. Gustafson 1994), hence PR drag is the dominant drag force around the Sun.

The timescale for a particle initially at $R_{\rm out}$ to be dragged to $R_{\rm in} \lesssim R_{\rm out}$ can be determined by integrating the secular (i.e., orbit-averaged) changes in the orbital parameters (see Equation (47) in J. A. Burns et al. 1979), which result in

$$t_{\rm prd} = \frac{c(R_{\rm out}^2 - R_{\rm in}^2)}{4\beta_{\rm rad}GM_*}, \tag{B12}$$

$$t_{\rm swd} = \frac{v_{\rm sw}(R_{\rm out}^2 - R_{\rm in}^2)}{4\beta_{\rm sw}GM_*}, \tag{B13}$$

$$t_{\rm combined} = t_{\rm prd} + t_{\rm swd} = \frac{R_{\rm out}^2 - R_{\rm in}^2}{4GM_*(\beta_{\rm rad}/c + \beta_{\rm sw}/v_{\rm sw})}. \tag{B14}$$

These equations show that the ratio of the drag timescales equals the ratio of the drag forces, i.e., the SW drag timescale is 10–20 times shorter than the PR drag timescales around $\epsilon$ Eri.

## ORCID iDs

Schuyler G. Wolff https://orcid.org/0000-0002-9977-8255
András Gáspár https://orcid.org/0000-0001-8612-3236
George Rieke https://orcid.org/0000-0003-2303-6519
Jarron M. Leisenring https://orcid.org/0000-0002-0834-6140
Antranik A. Sefilian https://orcid.org/0000-0003-4623-1165
Marie Ygouf https://orcid.org/0000-0001-7591-2731
Jorge Llop-Sayson https://orcid.org/0000-0002-3414-784X